\pdfoutput=1
\RequirePackage{rotating}
\documentclass[fleqn,usenatbib]{mnras}  

\usepackage[T1]{fontenc}
\usepackage{ae,aecompl}

\usepackage{newtxtext}
\usepackage[varvw]{newtxmath}

\usepackage{graphicx}	
\usepackage{amsmath}	
\usepackage[british]{babel}
\usepackage[]{units}
\usepackage{booktabs}
\usepackage{xcolor}
\usepackage{microtype}
\hypersetup{pdfauthor={Errani R., Navarro J.},
            pdftitle={The asymptotic Tidal Remnants of Cold Dark Matter Subhalos},
            pdfkeywords={},
            bookmarksnumbered=true}

\newcommand{\rmax}{r_\mathrm{mx}}	%
\newcommand{\rhomax}{\bar \rho_\mathrm{mx}}	%
\newcommand{\vmax}{V_\mathrm{mx}}	%
\newcommand{\Mmax}{M_\mathrm{mx}}	%
\newcommand{\tmax}{T_\mathrm{mx}}	%
\newcommand{\tasym}{T_\mathrm{asy}}	%
\newcommand{\kms}{\mathrm{km\,s^{-1}}}		%
\newcommand{\diff}{\mathrm{d}}
\newcommand{\Msol}{\mathrm{M_{\odot}}}

\newcommand{\kpc}{\mathrm{kpc}}
\newcommand{\Gyrs}{\mathrm{Gyrs}}
\newcommand{\rperi}{r_\mathrm{peri}}
\newcommand{\tperi}{T_\mathrm{peri}}
\newcommand{\Torb}{T_\mathrm{orb}}
\newcommand{\rapo}{r_\mathrm{apo}}
\newcommand{\maxzero}{_\mathrm{mx0}}
\newcommand{\rt}{{:}}

\title[Asymptotic tidal remnants of CDM subhalos]{The asymptotic tidal remnants of cold dark matter subhalos}

\author[]{Raphaël Errani$^{1,2,}$\thanks{errani@unistra.fr} \& Julio F. Navarro$^1$
\\
$^1$ Department of Physics and Astronomy, University of Victoria, Victoria, BC V8P 5C2, Canada\\
$^2$ Observatoire Astronomique, Université de Strasbourg, CNRS, 11 rue de l'Université, 67000 Strasbourg, France
}

\date{Accepted 2021 April 22. Received 2021 April 22; in original form 2020 November 16}

\pubyear{2020}
\volume{}

\begin{document}

\label{firstpage}
\pagerange{\pageref{firstpage}--\pageref{lastpage}} \pubyear{2020}
\maketitle

\begin{abstract}
  We use N-body simulations to study the evolution of cuspy cold dark
  matter (CDM) halos in the gravitational potential of a massive
  host. Tidal mass losses reshape CDM halos, leaving behind bound
  remnants whose characteristic densities are set by the mean density
  of the host at the pericentre of their respective orbit. The
  evolution to the final bound remnant state is essentially complete
  after $\sim 5$ orbits for nearly circular orbits, while reaching the
  same remnant requires, for the same pericentre, $\sim 25$
  and $\sim 40$ orbits for eccentric orbits with $1$:$5$ and $1$:$20$
  pericentre-to-apocentre ratios, respectively. The density profile of
  tidal remnants is fully specified by the fraction of mass lost, and
  approaches an exponentially-truncated Navarro-Frenk-White profile in
  the case of heavy mass loss. Resolving tidal remnants requires
  excellent numerical resolution; poorly resolved subhalos have
  systematically lower characteristic densities and are more easily
  disrupted. Even simulations with excellent spatial and time
  resolution fail when the final remnant is resolved with fewer than
  $3000$ particles. We derive a simple empirical model that describes
  the evolution of the mass and the density profile of the tidal
  remnant applicable to a wide range of orbital eccentricities and
  pericentric distances. Applied to the Milky Way, our results suggest
  that $10^8 $-$ 10^{10}\,\Msol$ halos accreted
  $\sim10\,\mathrm{Gyrs}$ ago on $1$:$10$ orbits with pericentric
  distance $\sim10\,\kpc$ should have been stripped to $0.1$-$1$ per
  cent of their original mass. This implies that estimates of the
  survival and structure of such halos (the possible hosts of
  ultra-faint Milky Way satellites) based on direct cosmological
  simulations may be subject to substantial revision.
\end{abstract}




\section{Introduction}
\label{sec:Introduction}

It is well established that structure in a universe dominated by cold dark matter (CDM) evolves hierarchically and leads to the formation of non-linear systems spanning an enormous range in mass \citep{WhiteRees1978,Frenk2012}. The basic units of this clustering hierarchy are CDM halos, virialized entities that form largely through the accretion, disruption, and merging of thousands of smaller subunits \citep[e.g.][and references therein]{Wang2020}. This complex merging process leaves behind an embedded population of ``subhalos''; i.e., the remnants of accreted subunits, many of which, despite shedding a large fraction of their initial mass, survive as recognizable self-bound entities for many orbital times \citep{Tormen1997,Ghigna1998,Klypin1999,Moore1999}.

It is now accepted that this halo {\it substructure} is a basic falsifiable prediction of the CDM paradigm, and underpins a number of observational efforts designed to probe the nature of dark matter on sub-galactic scales. Indeed, the role of substructure is critical to the interpretation of observational studies including, for example, (i) possible ``gaps'' in the tidal streams of disrupting globular clusters \citep[e.g.][]{Ibata2002,Johnston2002,ErkalBelokurov2014}; (ii) perturbations in strongly-lensed images of distant objects \citep[e.g.][]{Vegetti2009, Despali2017}; (iii) the number and long-term survival of faint satellite galaxies in the Galactic potential \citep[e.g.][]{Penarrubia2008b,Sanders2018,LiTucana2018};
and (iv) the ``boost factor'' of a potential dark matter annihilation signal \citep[e.g.][]{Tasitsiomi2002,DiemandKuhlenMadau2007,Lavalle2007,Springel2008b,Stref2019}.

Because of its complex origin, substructure in CDM halos is best studied via direct cosmological simulations, which have over the years converged on a basic outline of its basic properties. In the absence of baryons, for example, substructure is expected to be approximately self-similar, in the sense that the subhalo mass function, scaled to the host mass, rises steeply towards small masses and is similar for all virialized halos \citep{Kravtsov2004,Boylan-Kolchin2010,Wang2012,Jiang2016}. It is also widely accepted that substructure makes up only a small fraction ($\sim 5$-$10$ per cent) of the total mass of a halo, and that the subhalo spatial distribution and orbital properties are roughly independent of subhalo mass, especially at the low-mass end \citep{Springel2008a,Ludlow2009}.

Despite these advances, many substantive questions remain, especially those pertaining to the long-term survival of CDM subhalos and to the role of the central galaxy in aiding their tidal disruption \citep[e.g.][]{Johnston2002,Hayashi2003,DOnghia10,EPLG17,Garrison-Kimmel2017,vdBOgiya2018}. Also unclear is the final structure of heavily-stripped CDM subhalos, and the influence of numerical limitations on these results. These are important questions to resolve, as they may affect sensitively the theoretical interpretation of ongoing dark matter direct and indirect searches \citep[see, e.g.,][and references therein]{Green2005b}.

The issue of the long-term survival of CDM subhalos has been addressed in the past. While early work advocated for full subhalo disruption under certain conditions \citep[see; e.g.,][]{Hayashi2003}, more recent work has argued that, if the density profile of CDM halos is indeed cuspy  (i.e., $\diff \ln \rho / \diff \ln r = -1$ at the centre) as in  the Navarro-Frenk-White profile \citep[][hereafter NFW]{nfw1996,nfw1997}, then subhalos would rarely be fully disrupted and some form of bound remnant would almost always survive \citep[e.g.,][]{Penarrubia2010,vdb2018}. This is motivated by the fact that cuspy halos contain a substantial population of particles with extremely short orbital timescales \citep{EP2020}, which would always be ``adiabatically protected'' \citep{Weinberg1994} from the effects of tides.

Although there is growing consensus about this result, we note that it is unlikely to lead to a radical revision of the global properties of CDM substructure described above, which is dominated by subhalos affected only moderately by tidal effects. However, it may have important consequences for some detailed applications, especially those concerning substructure in the inner regions of a halo, where crossing times are short, where tides are most important, and where many observational studies focus on. 

A related issue is the structure of tidally-disrupted CDM subhalos, and, in particular, that of the final bound remnant, if indeed one survives. Prior work suggests that, as tides gradually truncate a subhalo, its characteristic parameters (i.e., radius, density, circular velocity) evolve along well-defined ``tidal tracks'' \citep{Penarrubia2008b}. There is, however, less consensus on how to describe the density profile of tidally-stripped subhalos; on how the final remnant properties depend on the strength of the tidal field; or on how long (i.e., number of orbits) it would take a subhalo to approach its asymptotic final state.

These are the issues we address here using idealized N-body simulations to follow the tidal loss/disruption of NFW halos in the potential of a massive host. The emphasis of our work is on the structure of the asymptotic tidal remnant of such halos, and on the timescale on which the process evolves.  This  paper is structured as follow: Sec.~\ref{sec:methods} introduces the numerical setup, including the host and subhalo models, as well as the initial conditions used in the simulations. The convergence of tidally stripped subhalos towards an asymptotic remnant is discussed in Sec.~\ref{sec:Convergence}, the effects of orbital eccentricity in Sec.~\ref{SecEcc}, while the tidal evolution of structural parameters and density profile shape are discussed in Sections~\ref{sec:Tracks} and \ref{sec:Profile}, respectively. The time evolution of bound remnants is discussed in Sec.~\ref{SecTimeEvol}. We describe simple applications of our modelling and compare with earlier work in Sec.~\ref{sec:Discussion}. We end with a brief summary of our main conclusions in Sec.~\ref{SecConc}. For completeness, numerical convergence issues are discussed in Appendix~\ref{SecNumConv}.

\section{Numerical methods}
\label{sec:methods}

We describe below the numerical setup of the simulations analyzed in this work. We assume, for simplicity, that the host halo may be approximated by a static, spherical potential, and that a CDM subhalo may be approximated by an NFW N-body model with mass much smaller than the host. We examine orbits that span a range of pericentric radii and eccentricities, and exercise care to monitor and exclude spurious results due to numerical limitations. 

\subsection{Host halo}
The host halo is represented by static, spherical isothermal potential,
\begin{equation}
  \Phi_\mathrm{host}(r) = V_0^2 ~ \ln\left( r/r_0 \right),
  \label{EqHalo}
\end{equation}
where $V_0=220\,\kms$ is the circular velocity and $r_0$ is an arbitrary reference radius.  The choice of a static, spherical potential ensures that the subhalo is subject to the same tidal field at each pericentric passage.
The corresponding circular velocity profile is flat and is chosen to match approximately the potential inferred for the Milky Way \citep[see e.g.][]{Eilers2019}.  The density profile is $\rho_\mathrm{host}(r) =\rho_0 (r/r_0)^{-2}$ (steeper than that of NFW halos at the centre; see Eq.~\ref{EqNFW}), with $V_0^2=4\pi G \rho_0 r_0^2$. These parameters correspond to a virial\footnote{We define the virial boundary of a halo as the radius where the mean enclosed density equals $200\times$ the critical density for closure, $\rho_{\rm crit}=3 H_0^2/8\pi G$, with $H_0 = \unit[67]{km\,s^{-1}\, Mpc^{-1}} $ \citep{Planck2018}. Virial quantities are denoted with ``200'' subscripts.} mass, $M_{200}=3.7 \times 10^{12} \, \Msol$, and a virial radius, $r_{200}=325$ kpc, at redshift $z=0$.

Although we quote below results for subhalos in solar masses, kpc, and km/s, these are only given for illustration and for ease of comparison with Milky Way subhalos. Gravitational effects are scale free, of course, and our results may be applied to any other value of $V_0$, or $r_0$, after proper scaling.

\subsection{Orbits}
\label{SecOrbits}

We explore tidal mass losses of subhalos on orbits with pericentre-to-apocentre ratios of $1\rt1$, $1\rt5$, $1\rt10$, and $1\rt20$. This eccentricity range includes those derived from Gaia proper motions for the orbits of (classical) Milky Way dwarf galaxies \citep{HelmiGaia2018,Fritz2018} and ultra-faint dwarfs \citep{Simon2018Gaia}.  All eccentric orbits are chosen to have an apocentric distance of $\rapo = 200\,\kpc$, and the subhalos are injected at apocentre.  The evolution of subhalos on circular orbits is studied as well, for orbital radii $r=40\,\kpc$ and $r=80\,\kpc$, respectively.

\subsection{N-body subhalos}
\label{SecSubhalos}

Subhaloes are modelled as N-body realizations of the NFW profile,
\begin{equation}
  \rho_{\rm NFW}(r) = {\rho_{\rm s} \over \left(r/r_\mathrm{s}\right) \left(1 + r/r_\mathrm{s}\right)^{2}},
  \label{EqNFW}
\end{equation}
where $r_\mathrm{s}$ is a scale radius and $\rho_{\rm s}$ is a characteristic density. The corresponding circular velocity of this profile peaks at $\vmax\approx1.65\,r_\mathrm{s}\,(G \rho_{\rm s})^{1/2}$ at a radius $\rmax \approx 2.16\, r_\mathrm{s}$. We shall adopt values measured at $\rmax$ as reference parameters in the analysis that follows. At that radius, the circular orbit time, $\tmax$, and characteristic mean enclosed density, $\rhomax$, may be written as 
\begin{equation}
\label{eq:tmaxtperi}
 \tmax = 2\pi \frac{\rmax}{\vmax} = \left( \frac{3\pi}{G \rhomax} \right)^{1/2}.
\end{equation}
Similarly, we define the mass $\Mmax \equiv M(<\rmax)$ enclosed within $\rmax$, and shall hereafter refer to $\rmax$, $\tmax$ and $\Mmax$ as the ``characteristic radius'', ``characteristic crossing time'' and ``characteristic mass'' of the subhalo, for short.

The NFW density profile has diverging total mass, so we exponentially truncate the profile outside $10\,r_\mathrm{s}$.
We generate isotropic, equilibrium models by sampling from the corresponding distribution function, obtained through Eddington inversion. We use the implementation described in \citet{EP2020}, which is available online\footnote{\url{https://github.com/rerrani/nbopy}}. 
Most of our realizations have $N=10^7$ particles, but we have varied this parameter extensively to check for numerical convergence. See App.~\ref{SecNumConv} for details on numerical convergence tests.

To limit the impact of orbital decay due to tidal mass losses (see, e.g., \citealp{White1983, HernquistWeinberg1989}, or more recently \citealp{Fuji2006,Fellhauer2007,Miller2020}), we choose an initial subhalo mass, $M\maxzero\equiv M(<r\maxzero) = 10^6\,\Msol$, much smaller than the host virial mass, and for which we have verified that the pericentric ($\rperi$) and apocentric ($\rapo$) distances do not change appreciably even after substantial tidal mass loss.

We are mainly interested in the regime where considerable tidal mass loss is expected, so we consider mainly cases where the initial characteristic density of a subhalo does not exceed the mean enclosed density of the host at pericenter. More precisely, we consider mainly cases where the initial characteristic crossing times, $T\maxzero$, compared with the circular time at pericentre, $\tperi = 2\pi \rperi /V_0$, satisfies $T\maxzero/\tperi \gtrsim 2/3$. We shall refer to this hereafter as the ``heavy mass loss regime''. We also report, for completeness, results for models with $ T\maxzero/\tperi \lesssim 2/3$ in Sec.~\ref{sec:AppendixB}. Overall, we have performed, for each orbit, simulations that span the range of characteristic crossing times, $0.2 < \tmax / \tperi < 2$.

\subsection{Particle-mesh and time integration}
\label{SecNumRes}

We follow the evolution of $N$-body subhalos in the tidal field of the host potential using the particle-mesh code \textsc{superbox} \citep{Fellhauer2000}. This code employs three cubic grids of $128^3$ cells each, two of them co-moving with the subhalo and centred on its centre of density. The highest-resolving co-moving grid has a resolution chosen to resolve the subhalo well, with grid size $\Delta x \approx r\maxzero / 128$, where $r\maxzero$ is the subhalo initial characteristic radius. The second co-moving grid has lower resolution, with grid size ten times larger, $\approx 10\,r\maxzero / 128$. The third grid has grid size $\approx 500\,\kpc/128$, is fixed in space, and is centred on the host potential. 

The time-integration is done using a leapfrog scheme with single and constant time step $\Delta t = \min( T\maxzero, \tperi )/400$. With these choices, a circular orbit at the finest grid resolution ($r \approx r\maxzero/128$) is resolved with (at least) $\approx 16$ time steps.

\subsection{Self-bound remnant}
This study focuses on the properties of self-bound dark matter substructures. We identify bound particles by (i) computing the centre of the subhalo via the shrinking sphere method \citep{Power2003}; (ii) computing the potential and kinetic energy of particles in a reference frame co-moving with the subhalo centre; (iii) discarding unbound particles in the co-moving frame; and iterating until convergence is reached or until the number of bound particles differs by less than one per cent from the previous iteration.

The properties of the self-bound remnant change abruptly as the subhalo passes through pericentre. Therefore, in what follows we choose to measure properties such as remnant density profiles, bound mass fractions, etc, at apocentre, where such properties are less subject to transient effects.

\begin{figure}
 \centering
 \includegraphics[width=\columnwidth]{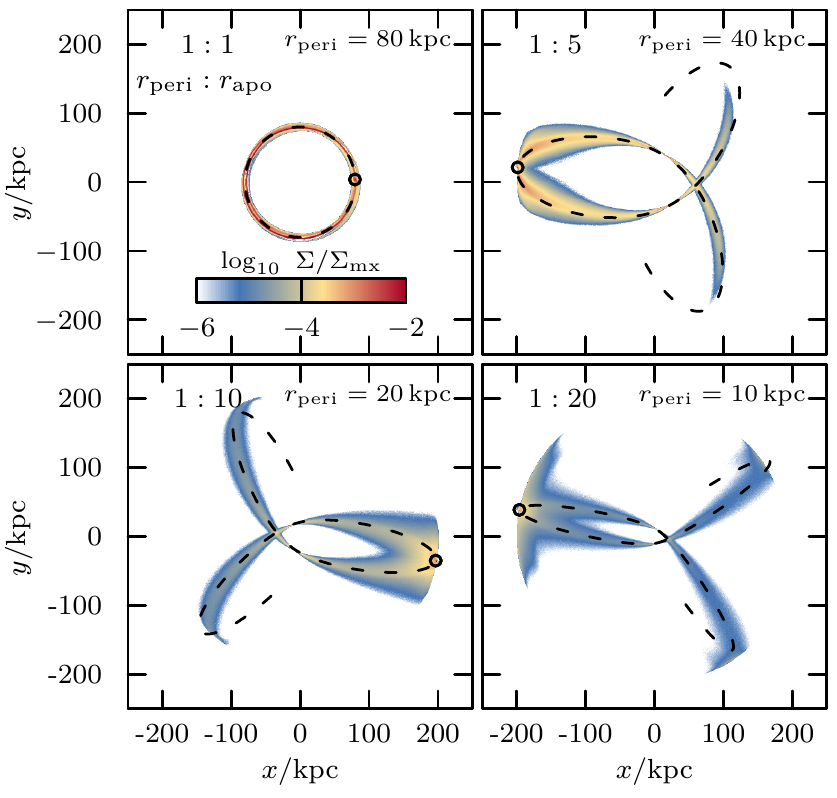}
 \caption{Tidal debris of CDM subhalos on four different orbits evolved in a spherical, isothermal potential (Eq.~\ref{EqHalo}). All subhalos have initial characteristic mass $M\maxzero = 10^6\,\Msol$, and crossing time $T\maxzero = 0.9\,\tperi$. The snapshots shown correspond to the $20^\mathrm{th}$ apocentric passage of each subhalo and show the debris on the orbital plane. The surviving bound remnant position is marked by an open circle. The projected density of tidally stripped material is colour coded, and normalized to the average projected density, $\Sigma_\mathrm{mx} = \Mmax/\pi \rmax^2$, of the bound remnant. The immediately preceding (and subsequent) orbital path of the remnant is shown by the dashed line in each panel.}
 \label{fig:orbit_overview}
\end{figure}

\begin{figure*}
 \centering
 \includegraphics[width=\textwidth]{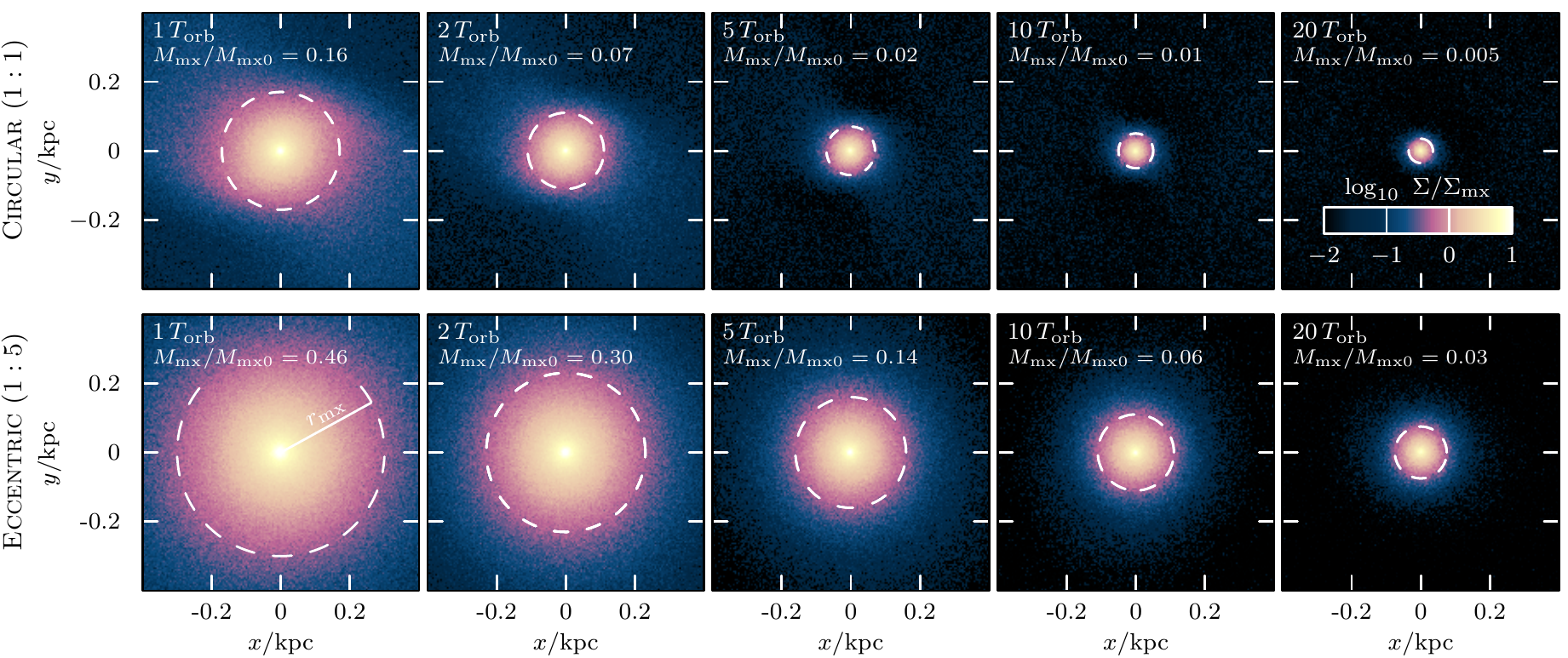}
 \caption{Projected structure of two bound remnants at various stages of their tidal evolution. The top row shows the projected densities of a subhalo on a circular orbit with $r=40\,\kpc$ in an isothermal potential after $1, 2, \dotsc, 20$ orbital periods.
 The subhalo has an initial characteristic mass $M\maxzero = 10^6\,\Msol$ and crossing time $T\maxzero = 0.88\,\tperi$, i.e. an initial characteristic radius and circular velocity of $r\maxzero=0.48\,\kpc$ and $V\maxzero=3.0\,\kms$, respectively. 
 The bound mass fraction is listed in the legend of each panel. The bottom row shows the same subhalo on an eccentric orbit with $r_\mathrm{peri} = 40\,\kpc$ and $r_\mathrm{apo} = 200\,\kpc$. It takes considerably longer to strip the subhalo on the eccentric orbit compared to the circular case. The projected density (normalized to the average projected density of the bound remnant, $\Sigma_\mathrm{mx} = \Mmax/\pi \rmax^2$) is colour-coded.}
 \label{fig:nbody_overview}
\end{figure*}

\begin{figure*}
 \centering
\includegraphics[width=8.5cm]{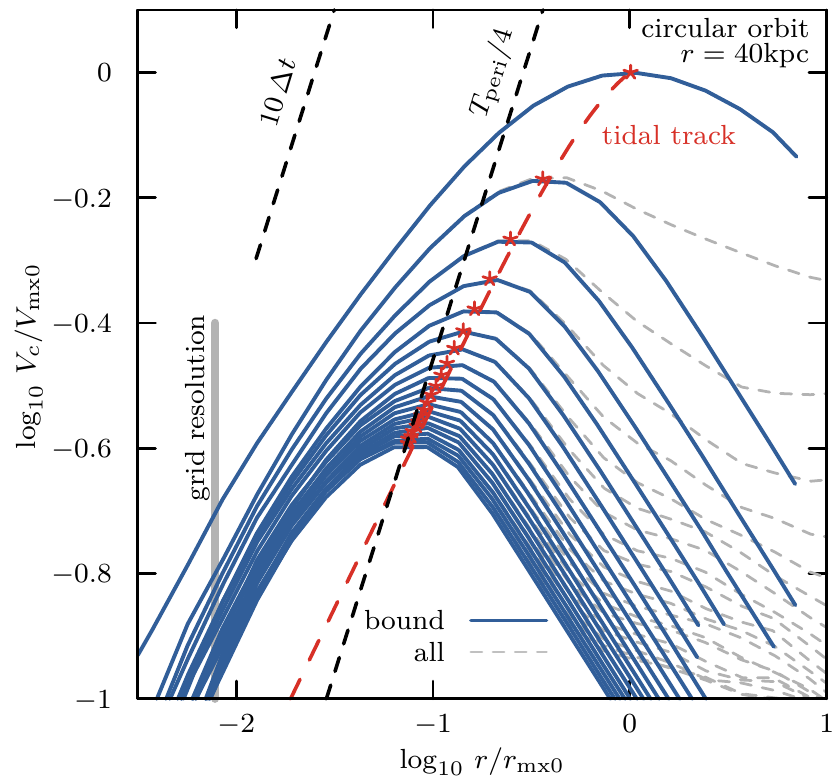}
\hspace*{0.4cm}
\includegraphics[width=8.5cm]{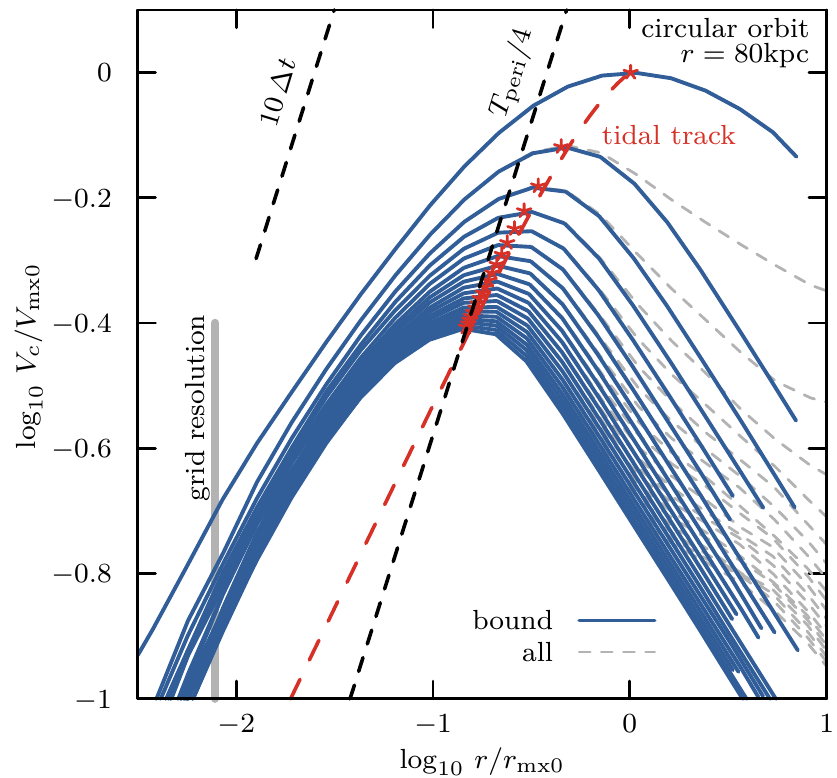}
\caption{Circular velocity profiles of the bound remnants of two subhalos on circular orbits, normalized to their initial characteristic radius and velocity. Grey curves correspond to all subhalo mass, blue curves to the self-bound remnant. Curves are spaced by one orbital period, for a total of 20 orbital periods. The left panel shows the evolution of the same subhalo as in the top panel of Fig.~\ref{fig:nbody_overview} ($r\maxzero=0.48\,\kpc$, $V\maxzero=3.0\,\kms$) on an orbit with $r=40\,\kpc$  
and $T\maxzero/\tperi=0.88$. The right panel shows a different subhalo ($r\maxzero=0.63\,\kpc$, $V\maxzero=2.6\,\kms$) on an orbit with $r=80\,\kpc$, which implies $T\maxzero/\tperi=2/3$ (for circular orbits $\tperi = \Torb$).
The evolution of $\{\rmax,\vmax\}$ follows well-defined \emph{tidal tracks} (red dashed curves), which we discuss further in Section~\ref{sec:Tracks}. In both cases, the evolution of the remnant slows down as $\tmax$ approaches $\tperi/4$. The final structure of a subhalo in the heavy mass-loss regime (i.e., $T\maxzero/\tperi>2/3$) is set solely by the properties of the host at pericentre. }
 \label{fig:vc_curves}
\end{figure*}

\section{Results}
\label{sec:results}

\subsection{General overview}

Fig.~\ref{fig:orbit_overview} shows the tidal debris of NFW subhalos placed on 4 different orbits of varying eccentricity and pericentric distance. The subhalos are shown at the 20th apocentric passage, with the immediately preceding (and following) orbital path indicated with dashed lines. The debris clearly stretches along the orbit, as expected for systems where the subhalo mass is negligible compared with the host. The colour scheme has been normalized to the maximum surface density of the bound remnant, which differs substantially from panel to panel because of the varying bound mass fraction of the remnant.

As expected, orbits with smaller pericentres lead to larger mass loss. This mass loss appears to continue as the subhalo continues to orbit the host, as shown in Fig.~\ref{fig:nbody_overview}, where the bound remnant of one subhalo is shown at various apocentric passages of the evolution for two orbits with the same pericentric distance, $\rperi = 40$ kpc. The top row corresponds to a circular orbit while the bottom row corresponds to an orbit with $1\rt5$ pericentre-to-apocentre ratio.

Fig.~\ref{fig:nbody_overview} illustrates a few interesting results. One is that, although for a given pericentre mass loss progresses faster in the case of a circular orbit (as expected), the remnant is qualitatively indistinguishable from that on the eccentric orbit after approximately the same mass fraction has been lost. Indeed, the circular orbit remnant after $2$ orbital periods looks similar to the $1\rt5$ orbit remnant after $10$ orbits; in both cases the bound remnant has retained roughly $6$-$7$ per cent of the initial $\Mmax$. Ditto for the  top-row remnant after $5$ orbits and the bottom-row remnant after $20$ orbits, when the bound remnant has been reduced in both cases to  $\sim 2$-$3$ per cent of the initial mass.

The second point to note from Fig.~\ref{fig:nbody_overview} is that, although mass loss is continuous, it slows down as the evolution progresses. For example, the eccentric-orbit subhalo takes only $\sim 1$ orbit to lose half of its mass at the beginning, but takes $\sim 10$ orbits to reduce its bound mass by the same factor between $10$ and $20\, T_{\rm orb}$. This suggests that a subhalo on an orbit with fixed pericenter is stripped until it converges to a well-defined self-bound ``asymptotic tidal remnant''. We explore this idea further below.

\subsection{Asymptotic tidal remnants}
\label{sec:Convergence}

\begin{figure}
 \centering
 \includegraphics[width=8.5cm]{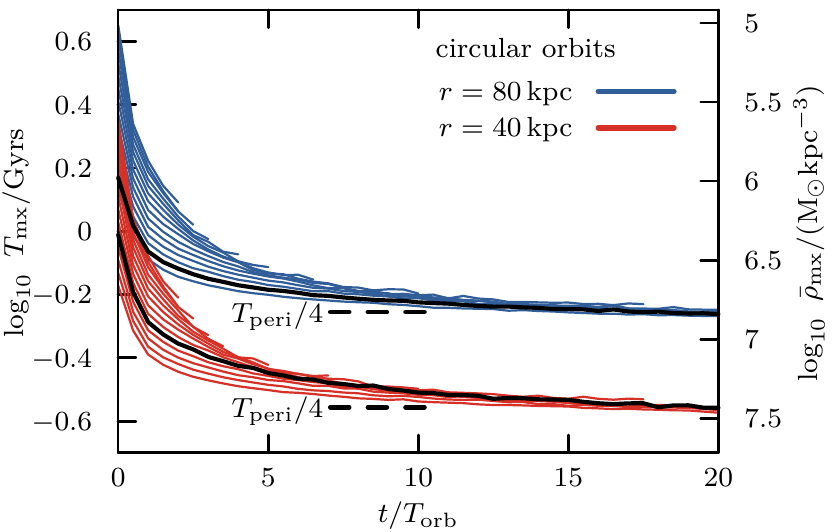}
 \caption{Evolution of the characteristic crossing time, $\tmax$, of subhalos on circular orbits with two different radii; $r=40$ kpc (red curves) and $r=80$ kpc (blue curves). Black curves highlight the two cases shown in Fig.~\ref{fig:vc_curves}. All subhalos are in the heavy mass-loss regime, with initial crossing times in the range $2/3 < T\maxzero/\tperi <2$, where $\tperi=\Torb$ for circular orbits. Subhalos are followed until their structure becomes compromised by numerical limitations, which become manifest when $\tmax$ has been reduced to less than $\sim1/3$ of its initial value for our $10^7$-particle realizations. See Appendix~\ref{sec:AppendixA} for further discussion on numerical convergence.  All subhalos are seen to approach an asymptotic value of $\tmax$ set solely by the host properties at the orbital pericentre.}
 \label{fig:tmax_convergence_circular}
\end{figure}

The effects of tidal mass loss are easily appreciated in Fig.~\ref{fig:vc_curves}, where we show the circular velocity profiles of two subhalos, placed on circular orbits with $r=40\,\kpc$ (left) and $r=80\,\kpc$ (right). The subhalo on the $40\,\kpc$ orbit is the one shown previously in the top panel of Fig.~\ref{fig:nbody_overview} with a ratio of crossing times of $T\maxzero/\tperi=0.88$, while the subhalo on the $80\,\kpc$ orbit has $T\maxzero/\tperi=2/3$. Curves are spaced by one orbital period, and each curve is normalized to the initial values of $\rmax$ and $\vmax$, which are $\{0.48\,\kpc,3.0\,\kms\}$ and $\{0.63\,\kpc,~2.6\,\kms\}$ for the subhalo on the $40\,\kpc$ and $80\,\kpc$ orbits, respectively.

The gradual convergence to a well-defined asymptotic remnant structure is quite clear; after $\sim 10$ orbits there is little further change in the mass profile of the remnant. The final characteristic density appears set by the mean density of the host at pericentre: more precisely, the subhalo is stripped gradually until its characteristic crossing time approaches a fixed fraction of the circular time at pericentre; $\tmax \approx \tperi / 4$, or, equivalently, until its characteristic density is $\sim 16\times$ the mean host density at pericentre. This is a general result of our simulations in the heavy mass-loss regime.

We illustrate this in Fig.~\ref{fig:tmax_convergence_circular}, which shows the evolution of the characteristic crossing time of subhalos in circular orbit at two different radii from the centre of the host: $40$ kpc (red) and $80$ (blue) kpc, respectively.  Each curve corresponds to subhalos with different initial characteristic densities, and follows a system for $20$ orbital times, or until its $\Mmax$ has been reduced to about $0.3$ per cent of its initial value, when numerical limitations begin to dominate (see App.~\ref{SecNumConv}). This mass reduction is equivalent to a reduction of nearly $\sim 16$ in the initial $\rmax$ or, alternatively, a factor of $\sim 5$ in $\vmax$ or $\sim 3$ in $\tmax$.

As is clear from Fig.~\ref{fig:tmax_convergence_circular}, all subhalos are stripped until their characteristic crossing times are reduced to $\tmax \approx \tperi / 4$, independent of the initial properties of the subhalo. This is true of all our runs in the ``heavy mass-loss regime'', where the initial characteristic density of the subhalo is low compared with the host density at pericenter (or, more precisely, when $T\maxzero / \tperi > 2/3$).

For comparison, we have computed characteristic crossing times of selected subhalos on circular orbits in the public DASH simulation suite \citep{Ogiya2019}, and observe that also there, tidal evolution decelerates, consistent with an evolution towards an asymptotic remnant.

\begin{figure}
 \centering
\includegraphics [width=\linewidth]{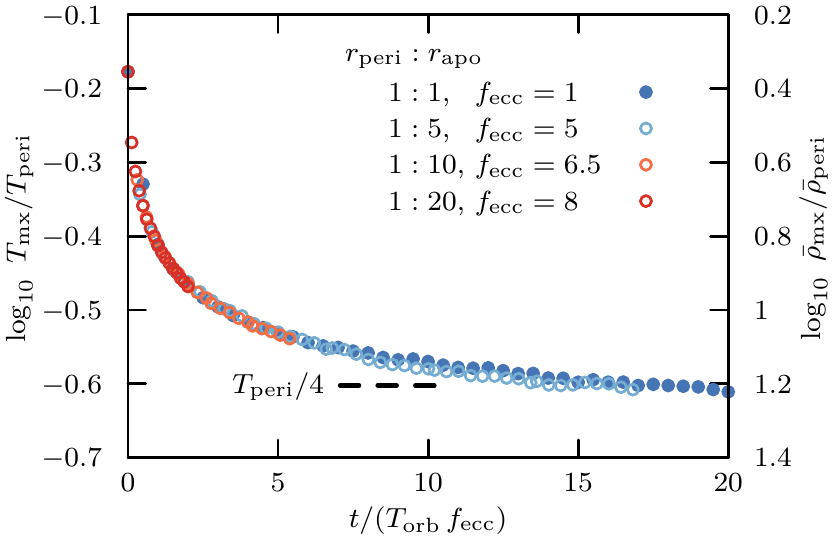}
\caption{Evolution of the characteristic crossing time, $\tmax$, (or, equivalently, of the characteristic density, scale on right) of a subhalo placed on orbits with fixed pericentric distance and varying eccentricities with pericentre-to-apocentre ratios 1:1, 1:5, 1:10, and 1:20. The evolution is similar in all cases, but occurs on longer timescales with increasing orbital eccentricity. At equal pericentre, the main effect of orbital eccentricity is to ``delay'' the tidal evolution of a subhalo. All subhalos evolve in the same way after scaling times by an eccentricity-dependent factor $f_{\rm ecc}$, listed in the legend. For example, it takes $f_{\rm ecc}=8$ times longer for a subhalo to be stripped to the same extent on a 1:20 orbit than on a circular orbit with the same pericentric radius.}
 \label{fig:tmax_time_eccentricities}
\end{figure}

\subsection{The effect of orbital eccentricity}
\label{SecEcc}

Circular orbits are rare in a cosmological setting, so it is important to explore how the results discussed above are modified for subhalos on eccentric orbits. As hinted at when discussing Fig.~\ref{fig:nbody_overview}, for given pericentre, tides are expected to operate on a longer timescale for eccentric orbits, mainly because tidal forces are strongest during pericentric passage and subhalos spend less time near pericentre the more eccentric the orbit. Is the tidal evolution on highly eccentric orbits just delayed, but otherwise similar to that on circular orbits?

We see that this is indeed the case in Fig.~\ref{fig:tmax_time_eccentricities}, where we show the evolution of $\tmax$ for a subhalo on four orbits with the same pericentre but different eccentricities. The subhalo has, initially, $T\maxzero=(2/3)\,\tperi$. The filled blue circles correspond to a subhalo that evolves on a circular orbit. Open circles correspond to results for  other orbital eccentricities, after scaling each in time by a factor, $f_{\rm ecc}=5$, $6.5$, and $8$ for orbits with pericentre-to-apocentre ratios of $1\rt5$, $1\rt10$, and $1\rt20$, respectively. The excellent agreement between the various curves confirms that the main effect of orbital eccentricity is simply a ``delay''.

In other words, it takes $5$ times more orbits for a subhalo on a $1\rt5$ orbit to evolve to the same stage as a subhalo on a circular orbit. Longer delays accompany higher eccentricities, but the delay factor appears to nearly saturate for eccentricities as high as $1\rt10$ or $1\rt20$, the highest value explored in our runs. We find that this is also a general result of our runs: all results obtained for circular orbits are generally applicable to other eccentricities simply by scaling time by the appropriate factor $f_{\rm ecc}$.

The following function may be used to interpolate between our four measured values of $f_{\rm ecc}$ for a given apocentre-to-pericentre ratio:
\begin{equation}
  f_{\rm ecc}\approx  \left[ 2x/(x+1)  \right]^{3.2}  ~~~  \mathrm{where}~ x = r_\mathrm{apo}/r_\mathrm{peri} ~.
  \label{EqFecc}
\end{equation}
A fitted exponent of $\approx 3.2$ reproduces the measured factors $f_{\rm ecc}$ at their respective ratios of $r_\mathrm{apo}/r_\mathrm{peri}$ within 5 per cent. 

Note that these factors are measured from simulated subhalos on orbits in an isothermal potential approximating the Milky Way (Eq.~\ref{EqHalo}). These factors may take slightly different numerical values in potentials with a substantially different radial dependence of the tidal forces.

We emphasize again that, while more pericentric passages are needed for a subhalo on an eccentric orbit to be tidally stripped to the same extent as on a circular orbit, the characteristic crossing time (density) of the asymptotic remnant (in the ``heavy mass loss regime'') is independent of orbital eccentricity and appears set solely by the circular time (density) of the host halo at pericentre.

\subsection{Tidal evolutionary tracks}
\label{sec:Tracks}

As may be seen in Fig.~\ref{fig:vc_curves}, the structural parameters $\rmax$ and $\vmax$  of the subhalos evolve along clearly defined ``paths'', indicated by the dashed red line in each panel. This is consistent with earlier work, which has shown that, as subhalos lose mass to tides, their characteristic parameters evolve along well-defined ``tidal tracks''. The position along the track depends only on the total amount of mass lost, and is largely independent of the eccentricity of the orbit and/or of the elapsed number of orbits. This was first discussed in \citealt{Penarrubia2008b}, hereafter P+08 (and confirmed in later work; see, e.g., \citealt{greenvdb2020}).

\begin{figure}
 \centering
\includegraphics [width=\linewidth]{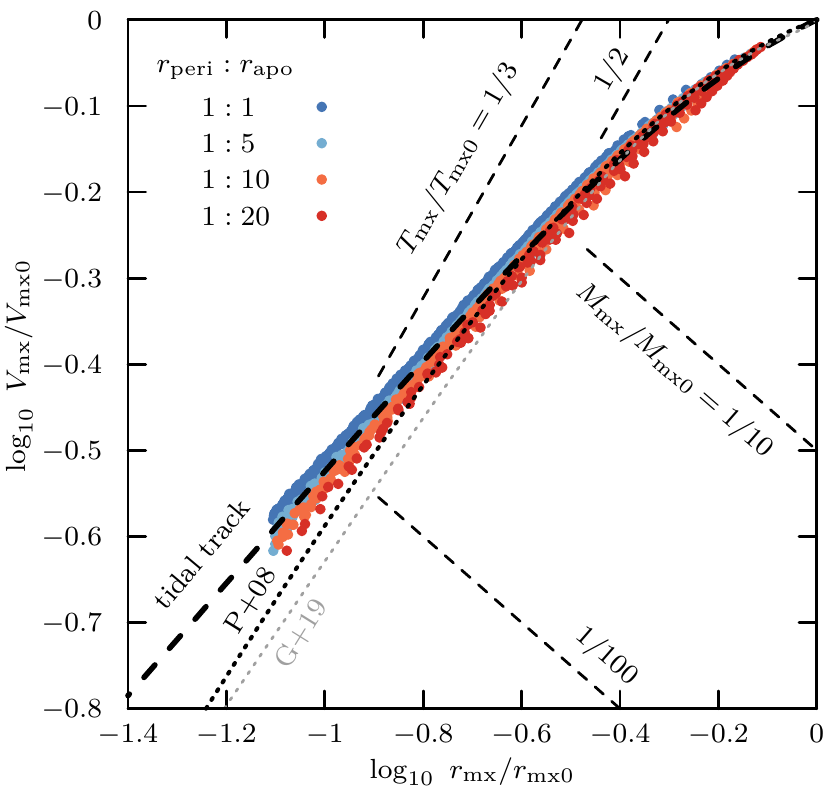}
\caption{Subhalo structural parameters ($\rmax,\vmax$, normalized to their initial values) evolve along a well-defined \emph{tidal track} that is nearly independent of orbital eccentricity. Filled circles correspond to measurements for {\it all} of our well-resolved N-body models, while the dashed black curve shows the fit given by Eq.~\ref{eq:tracks}. Constant remnant mass fractions, $\Mmax / M\maxzero$, and constant crossing time fractions, $\tmax / T\maxzero$, are shown by black dashed lines. The original tidal track of \citet{Penarrubia2008b} is shown by the black dotted curve, while the tracks of \citet{greenvdb2020} for subhalos of concentration $c = r_{200}/r_\mathrm{s} = 10$ are shown by a grey dotted curve. }
 \label{fig:all_tracks}
\end{figure}

We  explore this further in Fig.~\ref{fig:all_tracks}, where we show, for all of our runs, the evolution of the subhalo structural parameters $\{\rmax,\vmax\}$, normalized to their initial values $\{r\maxzero, V\maxzero\}$, and coloured by the eccentricity of the orbit. It is clear that a unique track describes well all runs, which may be parameterized by a simple function,
\begin{equation}
 \vmax / V\maxzero  = 2^\alpha~ \left(\rmax / r\maxzero\right)^\beta~ \left[1+(\rmax / r\maxzero)^2\right]^{-\alpha}~,
 \label{eq:tracks}
\end{equation}
with $\alpha = 0.4$, $\beta = 0.65$.  Note that this parametrization is slightly different from the one proposed by P+08 (shown with a black dotted line), an update made possible by the higher numerical resolution of our present runs, which give robust results for subhalos that retain as little as $0.3$ per cent of their initial characteristic mass, $\Mmax$.

An interesting feature of the tidal track is its clear curvature for modest mass losses  (i.e., for $\rmax/r\maxzero>1/3$, or $\Mmax/M\maxzero>0.1$) and a  power-law behaviour for heavier mass losses ($\Mmax/M\maxzero<0.1$), where the relation becomes
\begin{equation}
 \vmax/V\maxzero \propto (\rmax/r\maxzero)^{0.65},
\end{equation}
consistent with the power-law fits in \citet{EP2020}.
As we discuss below, the reason for this change is that heavily-stripped NFW halos converge to a new mass profile shape after substantial tidal mass loss. The curvature in the tidal track corresponds to the transition from the initial NFW mass profile to the new profile; once this is established further mass loss is ``self-similar'' and results in a simple power-law scaling between $\rmax$ and $\vmax$.

We emphasize again that the tidal track in Fig.~\ref{fig:all_tracks} applies equally well to {\it all} of our runs, regardless of pericentric radii and/or orbital eccentricity. This is true {\it provided} that the remnant can be adequately resolved.  As we discuss in App.~\ref{SecNumConv}, poor numerical resolution leads to systematic deviations from the tidal track, usually towards artificially low values of $\vmax$ and/or artificially large values of $\rmax$. These deviations result in characteristic crossing times longer than those of well-resolved subhalos, making the poorly resolved remnants prone to further tidal mass loss and eventual disruption. In what follows, we shall focus only on well-resolved systems, which we may define as those whose characteristic crossing times, $\tmax$, deviate by less than $10$ per cent from the tidal track given by Eq.~\ref{eq:tracks}. See App.~\ref{SecNumConv} for further discussion.

\begin{figure*}
 \centering
 \includegraphics[width=\textwidth]{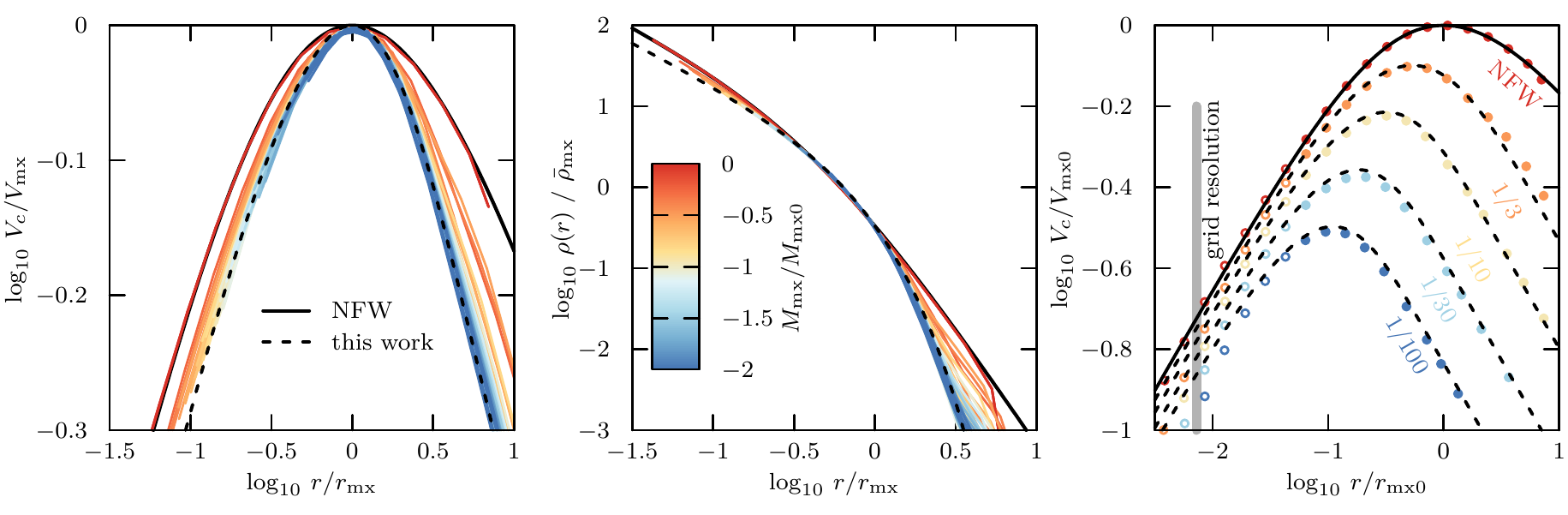} \caption{Mass profiles of stripped NFW subhalos. The left panel shows circular velocity profiles, normalized to their current values of $\rmax$ and $\vmax$, and coloured by the remaining self-bound mass fraction of the remnant (see colour bar in middle panel). The middle panel shows the density profiles of the same subhalos, normalized in a similar manner. Both of these panels show that the structure of a heavily-stripped NFW subhalo approaches a new, exponentially-truncated density profile whose shape is well approximated by Eq.~\ref{eq:rho_tidal}. The right-hand panel shows circular velocity curves for selected simulation snapshots at remnant bound masses of $\Mmax/M\maxzero = 1,\dotsc,1/100$ scaled to the initial values $\{ r\maxzero, V\maxzero \}$. Radii at which the circular velocity may be affected by resolution limitations according to the criteria of Appendix~\ref{sec:AppendixA} are shown using open circles, while filled circles correspond to radii unaffected by resolution. Exponentially truncated NFW profiles, with truncation radii $r_\mathrm{cut}$ selected to match the measured $\Mmax$ (see Fig.~\ref{fig:mmax_vs_rcut}), are shown using black dashed curves.}
\label{fig:profile_evolution}
\end{figure*}

\begin{figure}
\includegraphics[width=8.5cm]{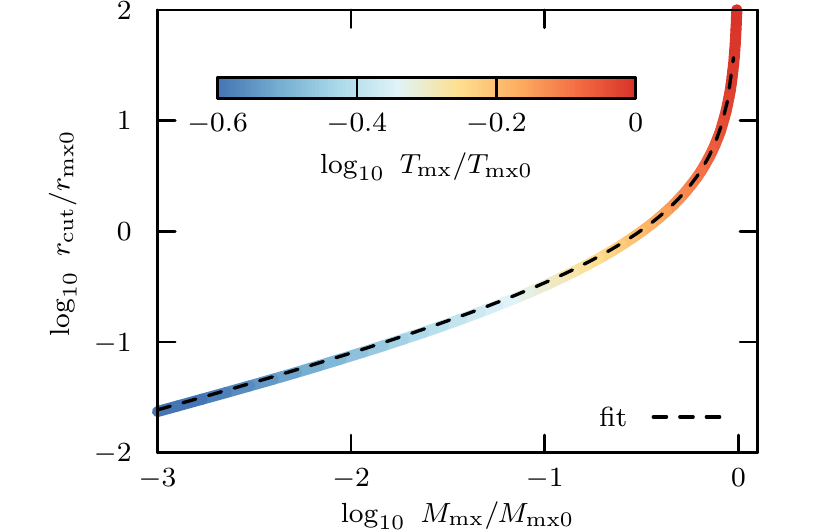}
 \caption{For the tidally truncated profile of Fig.~\ref{fig:profile_evolution} (see Eq.~\ref{eq:rho_tidal}), the truncation radius $r_\mathrm{cut}$ follows directly from the remnant bound mass fraction, $\Mmax/M\maxzero$. A truncation radius of $r_\mathrm{cut}/r\maxzero \rightarrow \infty$ recovers NFW, while for $r_\mathrm{cut}/r\maxzero \rightarrow 0$, the profile converges to an exponentially truncated cusp. While the relation of $\Mmax/M\maxzero$ and $r_\mathrm{cut}/r\maxzero$ follows directly from Eq.~\ref{eq:rho_tidal}, the dashed curve (Eq.~\ref{eq:rcutvsmmax}) provides a simple fit for ease of use.}
 \label{fig:mmax_vs_rcut}
\end{figure}

\subsection{Evolution of the density profile}
\label{sec:Profile}

As subhalos lose mass to tides, the shape of their mass profiles evolves from the original NFW shape adopted as initial conditions, and approaches a different profile shape after substantial mass loss has occurred. We show this in  Fig.~\ref{fig:profile_evolution}, where the left panel shows the circular velocity profiles of a number of subhalos on $1\rt5$ eccentric orbits, coloured by their remaining bound mass fraction, and scaled to their {\it current} values of $\rmax$ and $\vmax$. Similar results are obtained for all types of orbits; we choose here subhalos on $1\rt5$ orbits only as illustration.

The NFW profile is shown in Fig.~\ref{fig:profile_evolution} by the solid black line, and it agrees, by construction, with the initial subhalo profile (red curve). As a subhalo loses mass, the $V_c$ profile of its bound remnant becomes noticeably ``narrower'', with less mass in the outer regions, but also less mass in the regions inside $\rmax$ relative to the initial NFW profile. Gradually, this profile approaches a new asymptotic shape, which we indicate with the dashed black curve in Fig.~\ref{fig:profile_evolution}. 

The transition from the initial NFW density profile (Eq.~\ref{EqNFW}) to the asymptotic shape may be described by an exponential truncation of the initial profile, as follows:
\begin{equation}
 \rho(r)  = \rho_{\mathrm{NFW}}(r) ~ \times ~  {\exp( - r/r_\mathrm{cut} )}~/~{(1+ r_\mathrm{s}/r_\mathrm{cut} )^{\kappa}},
 \label{eq:rho_tidal}
\end{equation}
where $r_\mathrm{s}$ denotes the scale radius of the initial NFW profile, and $\kappa=0.3$ is chosen to match the ``tidal track'' evolution discussed above in Sec.~\ref{sec:Tracks}.
For $r_\mathrm{cut}/r_\mathrm{s} \rightarrow \infty$, this description recovers the initial NFW profile, whereas for $r_\mathrm{cut}/r_\mathrm{s}  \rightarrow 0$, the density profile converges to an exponentially truncated cusp.

For heavy mass losses, i.e. $r_\mathrm{cut}/r_\mathrm{s} \rightarrow 0$, equation \ref{eq:rho_tidal} reduces to an exponentially truncated cusp,
\begin{equation}
 \rho_\mathrm{asy}(r) = \rho_\mathrm{cut} ~ (r/r_\mathrm{cut})^{-1} ~ \exp( - r/r_\mathrm{cut} )~,
 \label{eq:rhoasy}
\end{equation}
where $\rho_\mathrm{cut} = \rho_\mathrm{s}~ (r_\mathrm{cut}/r_\mathrm{s})^{\kappa-1}$, and $r_\mathrm{s}$ and $\rho_\mathrm{s}$ denote the scale radius and scale density of the initial NFW profile, respectively. The asymptotic profile of Eq.~\ref{eq:rhoasy} has a convergent total mass of $M_\mathrm{tot} = 4 \pi r_\mathrm{cut}^3 \rho_\mathrm{cut}$,
and a circular velocity curve which peaks at a radius $\rmax \approx 1.8\,r_\mathrm{cut}$ with a characteristic mass of $\Mmax \approx 0.5\,M_\mathrm{tot}$. Consequently, this profile is consistent with the power-law tidal tracks $ \vmax \propto \rmax^\beta$, where $\beta = (1+\kappa)/2 \approx 0.65$ for a value of $\kappa \approx 0.3$.

For intermediate amounts of mass loss Eq.~\ref{eq:rho_tidal} describes well the profile of the remnant, with a value of $r_{\rm cut}/r_s$ that depends only on the current bound mass fraction.  While the relation of truncation radius $r_\mathrm{cut}$ and remnant bound mass $\Mmax$ follows directly from integrating Equation~\ref{eq:rho_tidal}, we present for ease of use the following fit, which reproduces well the relation shown in Fig.~\ref{fig:mmax_vs_rcut} (and is consistent with our simulations for the resolved range of remnant masses, $\Mmax/M\maxzero \gtrsim 1/300$):
\begin{equation}
 \frac{r_\mathrm{cut}}{r\maxzero} \approx 0.44 ~ \times ~\left(\frac{\Mmax}{M\maxzero}\right)^{0.44} ~ \left[1 - \left(\frac{\Mmax}{M\maxzero}\right)^{0.3} \right]^{-1.1}~.
 \label{eq:rcutvsmmax}
\end{equation}
Here $r\maxzero \approx 2.16\,r_\mathrm{s}$ is the characteristic radius of the initial NFW profile. The functional form of Eq.~\ref{eq:rcutvsmmax} ensures that (i) for $\Mmax/{M\maxzero} \rightarrow 1$, ${r_\mathrm{cut}}/{r\maxzero} \rightarrow \infty$ i.e. the profile prior to mass loss is an NFW profile, and (ii) for $\Mmax/{M\maxzero} \rightarrow 0$, the correct asymptotic bound mass of Eq.~\ref{eq:rhoasy} is recovered\footnote{In the asymptotic regime, integrating Eq.~\ref{eq:rhoasy} with $\rho_\mathrm{cut} = \rho_\mathrm{s}~ (r_\mathrm{cut}/r_\mathrm{s})^{\kappa-1}$ and $r_\mathrm{s} \approx r\maxzero/2.16$ yields ${r_\mathrm{cut}}/{r\maxzero}  \propto (\Mmax/{M\maxzero})^{1/(2+\kappa)}$ with exponent $1/(2+\kappa) \approx 0.44$.}.
The right-hand panel of Fig.~\ref{fig:profile_evolution} compares the results of this fitting formula with the profiles of simulated subhalos spanning two decades in mass loss, with excellent results.

\begin{figure}
\includegraphics[width=8.5cm]{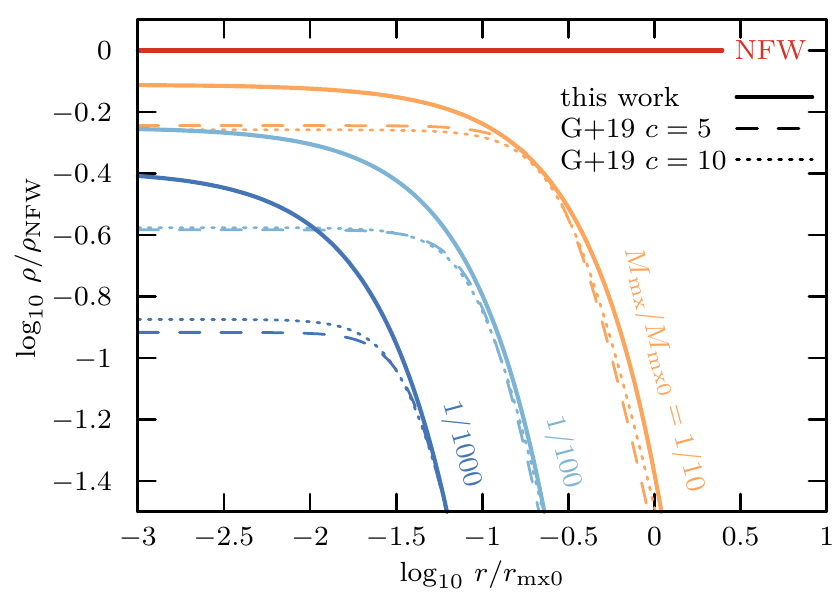}

\includegraphics[width=8.5cm]{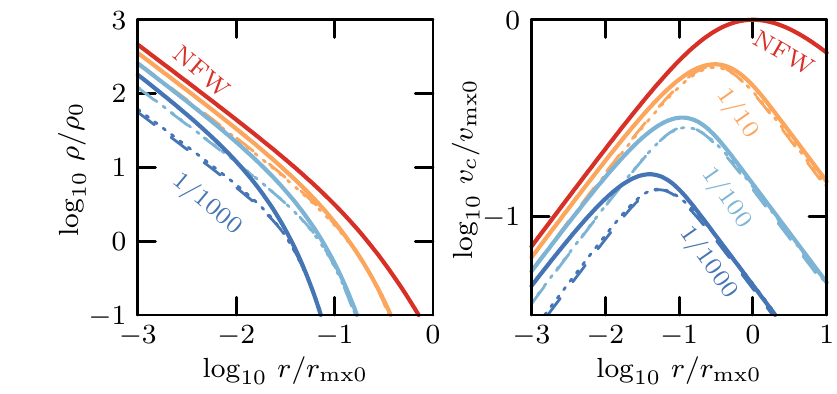}
 \caption{Transfer function $\rho/\rho_\mathrm{NFW}$ (top panel) as well as density profiles (bottom left) and circular velocity curves (bottom right) of the model of Eq.~\ref{eq:rho_tidal} for different fractions of bound mass $\Mmax / M\maxzero$. The model of Eq.~\ref{eq:rho_tidal} (``this work'', solid lines) is compared against the \citet{greenvdb2020} model for initial subhalo concentrations $c=r_{200}/r_\mathrm{s} = 5$ (``G+19'', dotted lines) and $10$ (dashed lines). Note that the model of Eq.~\ref{eq:rho_tidal} predicts higher central densities at equal bound mass fraction, and results in tidal tracks with a different asymptotic slope, see Fig.~\ref{fig:all_tracks}. }
 \label{fig:vdB_model}
\end{figure}

The model of Eq.~\ref{eq:rho_tidal} may be directly compared to that of \citet[][hereafter G+19]{greenvdb2020}, who propose a ``transfer function'', $\rho/\rho_\mathrm{NFW}$, to model the structural changes to NFW profiles during tidal evolution, fitted to simulation snapshots of the DASH simulation series \citep{Ogiya2019} . The transfer function corresponding to Eq.~\ref{eq:rho_tidal} is compared in Fig.~\ref{fig:vdB_model} to that of G+19. Note that our model leads to higher central densities at equal fractions of remnant bound mass $\Mmax/M\maxzero$. The main difference lies in the normalisation of the density profile, and not in its shape, as shown by the gradual divergence in the G+19 tidal track from ours seen in Fig.~\ref{fig:all_tracks}. Note, however, that even for the most highly stripped subhalo considered ``resolved'' in this work ($\Mmax/M\maxzero \sim 1/300$), the differences are rather small. Indeed, the G+19 track differs from ours there by less than $0.1$ dex in $\vmax$, or, equivalently, by less than $0.2$ dex in $\rmax$.

We turn our attention now to the time evolution of the characteristic parameters of the profile. Since the characteristic radius ($\rmax$) and velocity ($\vmax$) are linked by the tidal track shown in Fig.~\ref{fig:all_tracks} (Eq.~\ref{eq:tracks}) we only need to consider the evolution of one characteristic structural parameter to describe the full evolution. We choose the crossing time, $\tmax$, for this exercise next.

\subsection{Time evolution}
\label{SecTimeEvol}

We explore next how subhalos approach the asymptotic remnant stage as a function of time. This is illustrated in the top panel of Figure~\ref{fig:tmax_time_october}, which shows the evolution of $\tmax$ as a function of time for subhalos on circular orbits at $r=40$ kpc from the centre of the host. Times are scaled to the orbital time, $T_\mathrm{orb}$, and $\tmax$ is shown in units of the host circular time at pericentre, $\tperi$. Although, of course, $T_\mathrm{orb}=\tperi$ for circular orbits, this choice of scaling is useful, as it will enable us to extend the comparison to eccentric orbits, where the orbital time can be much longer than $\tperi$. We consider first only subhalos in the heavy mass-loss regime, i.e., $T\maxzero/\tperi>2/3$.

The top panel of Figure~\ref{fig:tmax_time_october} shows that all subhalos approach the same asymptotic remnant value, $\tasym$.  As discussed in Sec.~\ref{sec:Profile}, one may identify two phases in the evolution, one that applies to early times, when the subhalo mass profile shape changes rapidly from NFW-like to a new shape, and another one as all subhalos approach the same asymptotic remnant stage. During the first stage subhalos with larger values of $T\maxzero/\tperi$ evolve more rapidly, but they all seem to approach the same asymptotic behaviour after roughly $\sim 10$ orbits.

\subsubsection{Heavy mass loss regime: late asymptotic behaviour}

\begin{figure}
  \centering
\includegraphics[width=8.5cm]{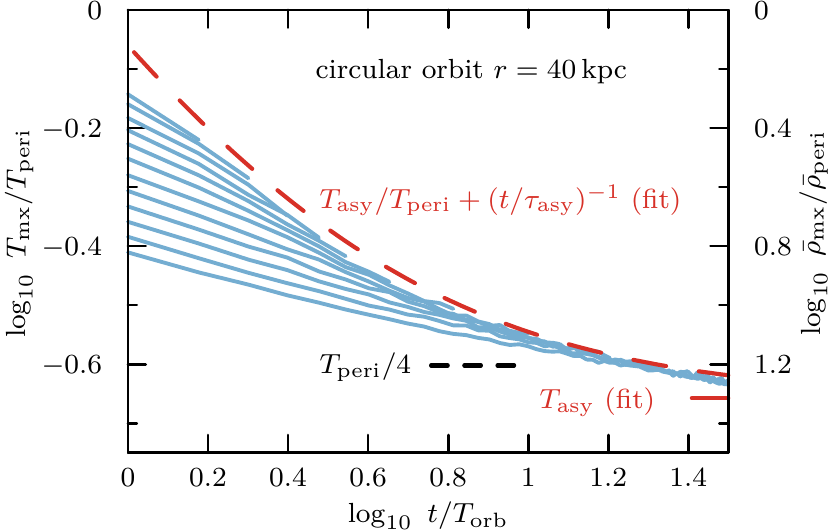}

\vspace*{0.3cm}

\includegraphics[width=8.5cm]{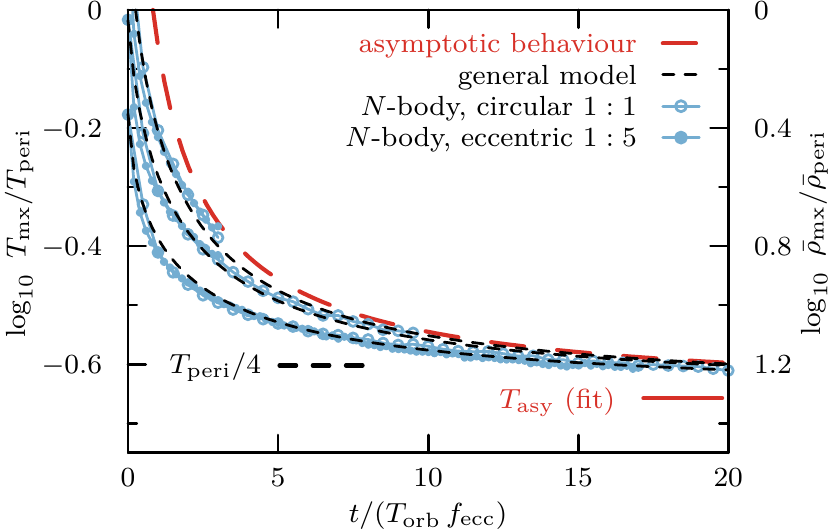}
\caption{{\it Top:} Time evolution of the characteristic crossing time of subhalos on circular orbits in the heavy mass-loss regime. The evolution is shown  in logarithmic units for a number of subhalos with initial crossing times of $  T\maxzero/\tperi > 2/3 $ in order to highlight the long-term behaviour of the subhalos, all of which approach at late times the same asymptotic power-law trend, highlighted by the red dashed curve and parametrized by Eq.~\ref{EqAsympt}. Short horizontal segments indicate the value of $\tmax$ to which bound remnants converge in our simulations, $\tperi/4$, which differs slightly from the fitted ``true asymptotic'' value (i.e., applicable for $t=\infty$), $\tasym \approx 0.22\,\tperi$. {\it Bottom:} Results for three subhalos of different initial, $T\maxzero/\tperi$, placed on different orbits and compared with the empirical model parametrized by Eq.~\ref{EqYevol}. Circular orbits are shown with open circles; filled circles correspond to $1\rt5$ eccentric orbits with the same pericentric radius. Eccentric orbit times have been scaled by $f_{\rm ecc}=5$, as discussed in Sec.~\ref{SecEcc}. Aside from this delay, Eq.~\ref{EqYevol} describes very well the results of all simulations.}
\label{fig:tmax_time_october}
\end{figure}

The asymptotic behaviour may be approximated by a simple power law (red dashed line in Figure~\ref{fig:tmax_time_october}),
\begin{equation}
  Y_{\rm asy}(t) \equiv  ({\tmax(t) - \tasym})/{\tperi} =  (t/\tau_{\rm asy})^{-1}.
  \label{EqAsympt}
\end{equation}
With $\tasym \approx 0.22\,\tperi$ and $\tau_\mathrm{asy} \approx 0.65\,\Torb$, this equation describes well the late stages of all of our runs in the heavy mass-loss regime. This power-law is the solution to the differential equation
\begin{equation}
  \label{eq:late_evol}
 \diff  Y_{\rm asy}(t) / \diff t = - Y_{\rm asy}^2(t) / \tau_{\rm asy},
\end{equation}
hence the asymptotic evolution of a subhalo's crossing time is such that the slope $ \diff  Y_{\rm asy}(t) / \diff t$ depends on the instantaneous value of $Y_{\rm asy}(t)$ alone, consistent with the observation that the subhalo profile shape converges: once the profile shape has converged, the subhalo structure is fully determined by the single parameter $\tmax(t)$. This late evolution may be thought of as ``self-similar'' in the sense that it is independent of the initial conditions, and progresses at a rate governed only by the instantaneous value of $\tmax - \tasym$.

\subsubsection{Heavy mass loss regime: general description}
\label{sec:general_model}

The early evolution deviates from the asymptotic power-law behaviour discussed above. The following empirical formula describes well the general evolution in the heavy mass-loss regime:
\begin{equation}
Y(t) = ({\tmax(t) - \tasym})/{\tperi} = Y_0 \left[ 1 + (t/\tau \right)^\eta]^{-1/\eta}
\label{EqYevol}
\end{equation}
where $Y_0 = (T\maxzero - \tasym)/\tperi$ is determined by the initial conditions, and $\eta$ is a free parameter that may be inferred from the simulation results.

With this parametrization, the fact that all subhalos in Figure~\ref{fig:tmax_time_october} approach the {\it same} late evolution implies that the timescale $\tau$ is inversely proportional to $Y_0$, i.e., 
\begin{equation}
 \tau = \tau_\mathrm{asy} /Y_0.
\end{equation}
Least-squares fits to the simulation data show that $\eta \lesssim 1$, and that $\eta$ correlates with $Y_0$, within ten per cent of the following empirical function:
\begin{equation}
\eta \approx 1 - \exp(-2.5 \, Y_0 ) ~.
\end{equation}
The bottom panel of Figure~\ref{fig:tmax_time_october} compares fits using Eq.~\ref{EqYevol} with simulation results. For clarity, we show only three different subhalos on circular orbits (open circles), but include also the results for orbits with $1\rt5$ pericentre-to-apocentre ratio (filled circles). As expected from our discussion in Sec~\ref{SecEcc}, open and filled circles overlap after the eccentric orbital times are scaled by $f_{\rm ecc}=5$. Aside from this eccentricity-dependent ``delay'', Eq.~\ref{EqYevol} describes well the overall evolution of all runs, regardless of orbital eccentricity.

A simple implementation of this model for the tidal evolution of subhalos is made available online\footnote{\url{https://github.com/rerrani/tipy}}. The implementation takes as inputs the initial subhalo structural parameters $\{r\maxzero,V\maxzero\}$ as well as host halo crossing time $\tperi$ at pericentre, orbital period $T_\mathrm{orb}$ and pericentre-to-apocentre ratio, and returns the time evolution of the subhalo structural parameters $\{\rmax(t),\vmax(t) \}$.

\subsubsection{Modest mass loss regime}
\label{sec:AppendixB}

\begin{figure}
 \centering
\includegraphics[width=8.5cm]{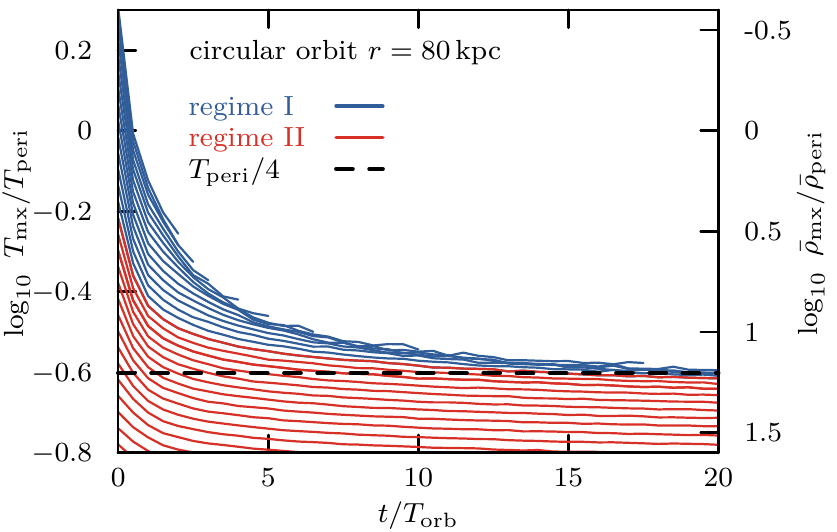}
\caption{As Fig.~\ref{fig:tmax_time_october}, but for circular orbits with $r=80$ kpc, and spanning a wide range of $T\maxzero/\tperi$. Halos in the heavy mass-loss ``regime I'' (i.e., $T\maxzero/\tperi>2/3$) are shown in blue, those in the modest mass-loss ``regime II'' (i.e., $T\maxzero/\tperi<2/3$) are shown in red. In regime I all halos converge to remnants with the same asymptotic value of $\tmax\approx \tperi/4$. In regime II, halos approach a remnant whose characteristic crossing time (density) depends on their initial value. Most subhalos in a cosmological context fall in regime~I (see Sec.~\ref{sec:Convergence} for details).  }
 \label{fig:tmax_time_dense_new}
\end{figure}

\begin{figure}
 \centering
 \includegraphics[width=8.5cm]{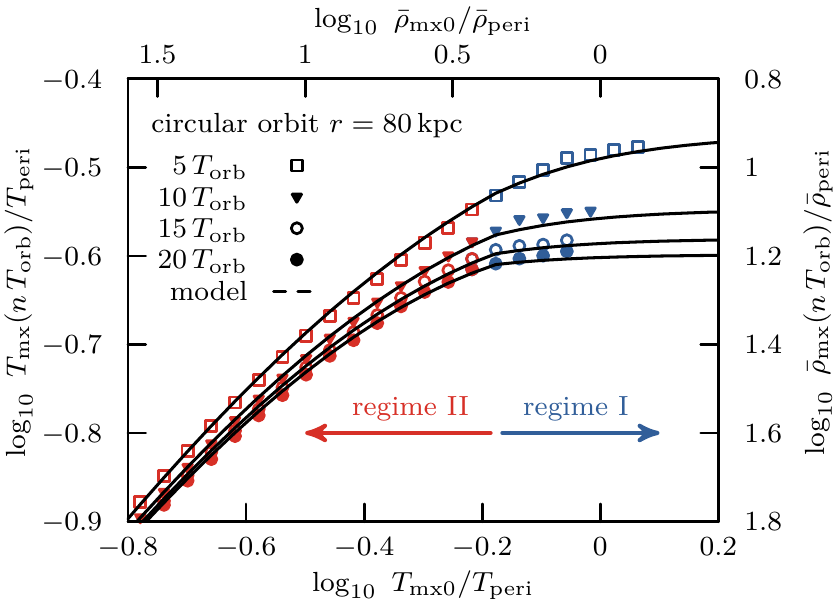} \caption{Crossing times of subhalos after $n=5$, $10$, $15$, and $20$ orbital times. The characteristic densities of subhalos in regime I (i.e., $T\maxzero/\tperi>2/3$) converge to an asymptotic value that is set solely by $\tperi$ and is independent of the initial value $T\maxzero$. Those in regime II (i.e., $T\maxzero/\tperi<2/3$) converge to characteristic densities that reflect their initial values. Solid black curves show the empirical results from Eq.~\ref{eq:model-regime-II-new} for regime II, and from Eq.~\ref{EqYevol} for regime I. }
 \label{fig:tmax_time_convergence_new}
\end{figure}

Subhalos with characteristic densities substantially higher than the host density at the pericentre of their orbits will be only modestly affected by tides. In this regime (i.e., when $T\maxzero/\tperi<2/3$), the remnant is {\it not} expected to have the same characteristic density as the asymptotic tidal remnant discussed in the preceding subsection. Their characteristic densities must somehow in this case reflect their initial values.

Fig.~\ref{fig:tmax_time_dense_new} shows the evolution of $\tmax$ for subhalos on circular orbits with $r=80\,\kpc$, and $0.2 < T\maxzero/\tperi < 2$. The evolution of subhalos with $T\maxzero/\tperi > 2/3$ (i.e., in the heavy mass-loss regime or ``regime I'', shown with blue curves) are analogous to those discussed above, and are seen to approach remnants with the same asymptotic crossing time, $\sim \tperi/4$.

On the other hand, subhaloes with $T\maxzero/\tperi < 2/3$ are shown using red curves. Tidal effects on these halos are modest, and the evolution of $\tmax$ quickly stalls after a few orbits. After $20$ full circular orbital periods the remnants have not yet settled to a final value, but evolve only weakly thereafter.

We may fit the tidal evolution of these subhalos  using the same  Eq.~\ref{EqYevol}, with ``primes'' to distinguish parameters specific to the modest mass loss regime (``regime II''): 
\begin{equation}
\label{eq:model-regime-II-new}
Y'(t) = Y'_{0} \left[ 1 + (t/\tau' \right)^{\eta'}]^{-1/\eta'} ,\hfill 
\end{equation}
where $Y' = (\tmax - \tasym')/\tperi$. The exponent $ \eta' = 0.67 $ may be fixed by requiring that it should be
identical to the exponent of Eq.~\ref{EqYevol} at the boundary between regimes I and II.
The main difference from the previous results is that, in regime II, the ``asymptotic'' crossing time $\tasym'$ depends on the initial $T\maxzero$ of the subhalo, and not solely on $\tperi$. We estimate $\tasym'$ through the following empirical function, 
\begin{equation}
 \label{eq:tasym_regime2}
 \tasym' /\tperi=  \frac{T\maxzero/\tperi}{(1+T\maxzero/\tperi)^\gamma } ~, \hfill \text{(regime II)}
\end{equation}
where the functional form is motivated by the crossing time dependence on initial conditions shown in Fig~\ref{fig:tmax_time_convergence_new}, discussed below. A choice of $\gamma\approx2.2$ ensures that at the boundary between regime I and II, the fitted asymptote $\tasym = 0.22\, \tperi$ of regime I is matched.
Using these constraints, the fitted decay rate $\tau'$ correlates with the initial crossing time $T\maxzero/\tperi$ roughly as
\begin{equation}
\tau' / \Torb= 1.2  (T\maxzero/\tperi)^{-1/2} ~.\hfill 
\end{equation}
While these parameters were determined for circular orbits, the extension to eccentric orbits is straightforward through the delay factor $f_\mathrm{ecc}$ discussed in Sec.~\ref{SecEcc}.

Figure~\ref{fig:tmax_time_convergence_new} compares $\tmax$ measured from $N$-body snapshots after $n=5,10,15$ and $20$ orbital periods for different initial crossing times $T\maxzero/\tperi$ against the empirical results of Eq.~\ref{eq:model-regime-II-new}, showing good agreement between the model and the simulations.
The functional dependence of the (near) asymptotic crossing time after $n=20$ orbital periods on initial conditions is well described by a function of the form of Eq.~\ref{eq:tasym_regime2}, which imposes that for $T\maxzero/\tperi \rightarrow 0$,  $\tasym \rightarrow T\maxzero$. In the regime where the subhalo is significantly denser than the host halo at pericentre, tidal evolution becomes negligible, as expected.

\begin{figure*}
 \centering
\includegraphics[width=8.5cm]{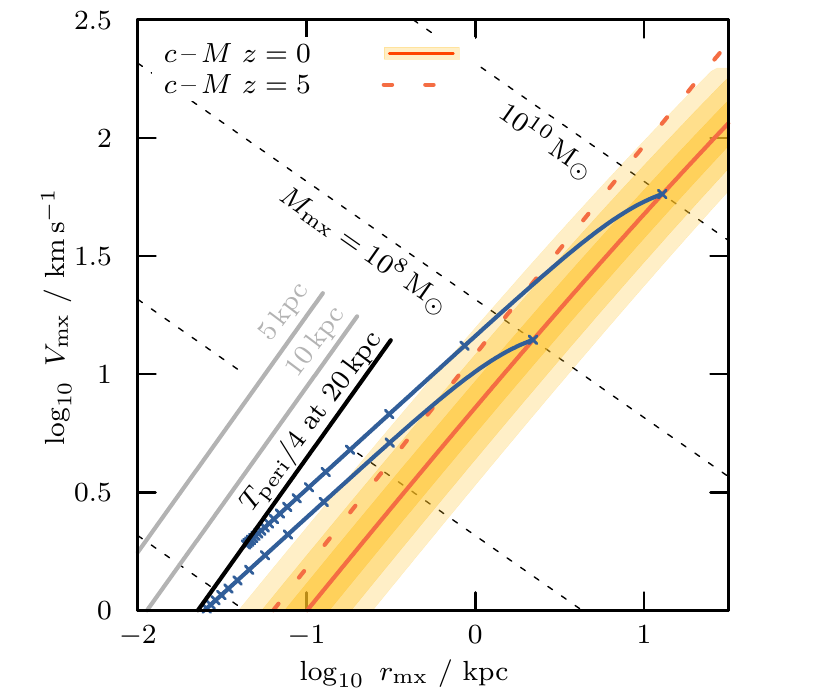}
\hspace*{0.3cm}
\includegraphics[width=8.5cm]{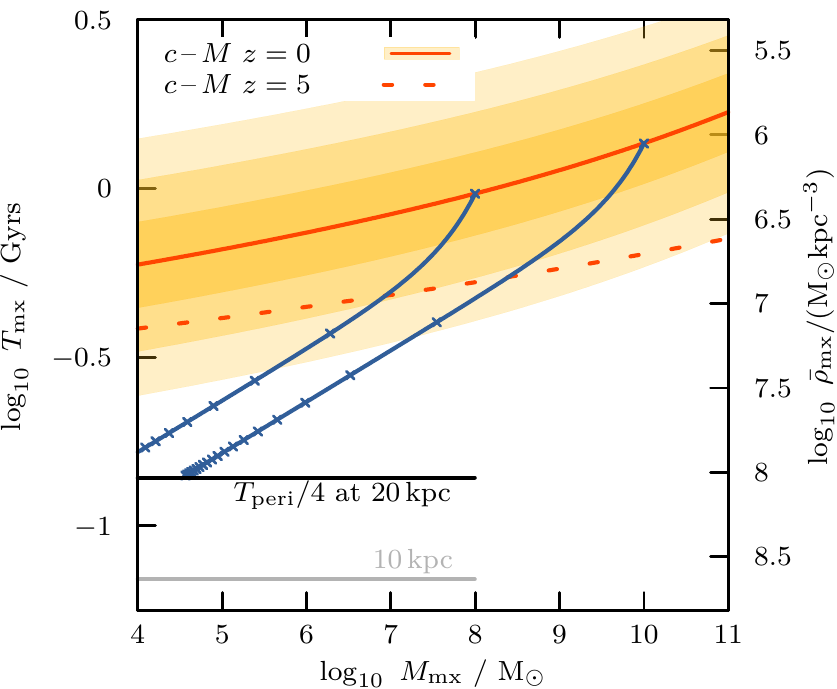}
 \caption{Tidal evolution in a cosmological context. The left panel shows $\{\rmax,\vmax\}$ tidal evolutionary tracks (blue solid curves) for subhalos of bound mass $\Mmax = 10^8\,\Msol$ and $10^{10}\,\Msol$ (i.e. virial masses and concentrations of $M_{200 } = 3.8 \times 10^8\,\Msol $, $c = r_{200}/r_\mathrm{s} \approx 15$  and $3.5 \times 10^{10}\,\Msol$, $c\approx11$, respectively.) with initial conditions consistent with the mean $z=0$ mass-concentration relation from \citet[][]{Ludlow2014} (yellow shaded bands, for successive $\pm0.1\,\mathrm{dex}$ scatter in concentration). Ticks along the tracks correspond to intervals of one orbital period $T_\mathrm{orb}$ for a circular orbit with $r = 20 \kpc$ in an isothermal potential with circular velocity $220\, \kms$ (or equivalently to $5\,T_\mathrm{orb} $ for $1\rt5$ orbits, and $6.5\,T_\mathrm{orb}$ for $1\rt10$ orbits with the same pericentre). For reference, the crossing times of asymptotic tidal remnants, $\tperi/4$, is shown for values of $r_\mathrm{peri}=5\,\kpc$, $10\,\kpc$, $20\,\kpc$.  The panel on the right shows the same tidal tracks, but in terms of bound mass, $\Mmax$, and crossing time, $\tmax$.   }
 \label{fig:rmaxvmax_mmaxtmax}
\end{figure*}

\section{Discussion}
\label{sec:Discussion}

The results of the previous section may be used to provide some insight into ongoing discussions regarding substructure in CDM halos and, in particular, on the abundance, structure, and spatial distribution of tidally-stripped subhalos. As discussed in Sec.~\ref{sec:Introduction}, these discussions concern a wide variety of topics, from the ultimate survival of dark matter dominated systems, such as faint satellite galaxies, to the interpretation of distortions of strongly-lensed galaxies, to theoretical expectations for a possible annihilation signal from surviving subhalos. We plan to address some of these in future contributions, but provide here a first application to a few topical issues as illustration.

\subsection{Tidal remnants in Milky Way-like systems}

Our discussion so far has dealt with subhalos with arbitrary values of $\rmax$ and $\vmax$, but these parameters are expected to be strongly correlated because of the redshift-dependent $\Lambda$CDM mass-concentration relation \citep[see; e.g.,][and references therein]{Ludlow2014}. This is shown, for illustration, in the left panel of Fig.~\ref{fig:rmaxvmax_mmaxtmax}, where the solid red line indicates the mean relation at $z=0$ and the shaded bands correspond to succesive $\pm 0.1\,\mathrm{dex}$ scatter in concentration. We also indicate, for completeness, the expected mean relation at $z=5$ with a dashed red line.

$\Lambda$CDM subhalos are constrained to move along the tidal track discussed in Sec.~\ref{sec:Tracks}, two examples of which are shown by the blue curves in Fig.~\ref{fig:rmaxvmax_mmaxtmax}. One of them corresponds to a halo with initial $\Mmax=10^{10}\, M_\odot$ and the other to $\Mmax=10^{8}\, M_\odot$.
Assuming that these halos were placed on circular orbits in a potential like that of the Milky Way (represented crudely by Eq.~\ref{EqHalo}) at $r=20$ kpc, these subhalos would be quickly stripped of mass (each tickmark on the tracks corresponds to one orbital period), and would gradually approach the asymptotic remnant stage, where $\tmax \approx \tperi/4$ (shown by the thick black line). We see from this that a $10^{10}\, M_\odot$ halo would leave behind an asymptotic remnant with less than $10^5\, M_\odot$, a characteristic radius of $\rmax\sim 30$ pc and a maximum circular velocity of  $\vmax\sim 2$ km/s.

Such remnants are essentially impossible to properly resolve in direct cosmological simulations; indeed, a $10^{10}\, M_\odot$ subhalo would be resolved with fewer than $\sim 10^6$ particles in even some of the highest resolution simulations ever completed, such as those from the Aquarius project \citep{Springel2008b}. As discussed in Appendix~\ref{sec:AppendixA}, a subhalo with $10^6$ particles starts to deviate from the correct tidal track after being reduced to about than $1/100$ of its initial mass, becoming increasingly prone to full (and artificial) tidal disruption. This implies that essentially no surviving $10^5\, M_\odot$ halos would be direct descendants of systems with initial mass of order $10^{10}\, M_\odot$, as such systems would be most likely fully disrupted.

We note that this does not mean that the abundance of surviving $10^5\, M_\odot$ halos has been severely underestimated in simulations like Aquarius. Indeed, the abundance of low-mass subhalos is 
vastly dominated by recently accreted low mass subhalos that have been only modestly stripped; in other words, there are simply too few $10^{10}\, M_\odot$ systems to change the abundance of $10^5\, M_\odot$ subhalos much \citep{Springel2008b}.

We also note that the comments above refer to the asymptotic tidal remnant of a subhalo, which is only reached after completing a fairly large number of orbits. In reality, most subhalos have only had time to complete a few orbits, depending on their accretion time and their apocentric distance. We may use the time evolution model described in Sec.~\ref{SecTimeEvol} to take this into account and to estimate the present-day mass of subhalos accreted at different times during the evolution of a Milky Way-like halo. Since our main goal is to illustrate possible applications of our results, rather than to provide detailed predictions, we shall assume for this exercise that the host halo remains unchanged throughout and that it is well approximated by Eq.~\ref{EqHalo}.

\begin{figure}
 \centering
\includegraphics[width=8.5cm]{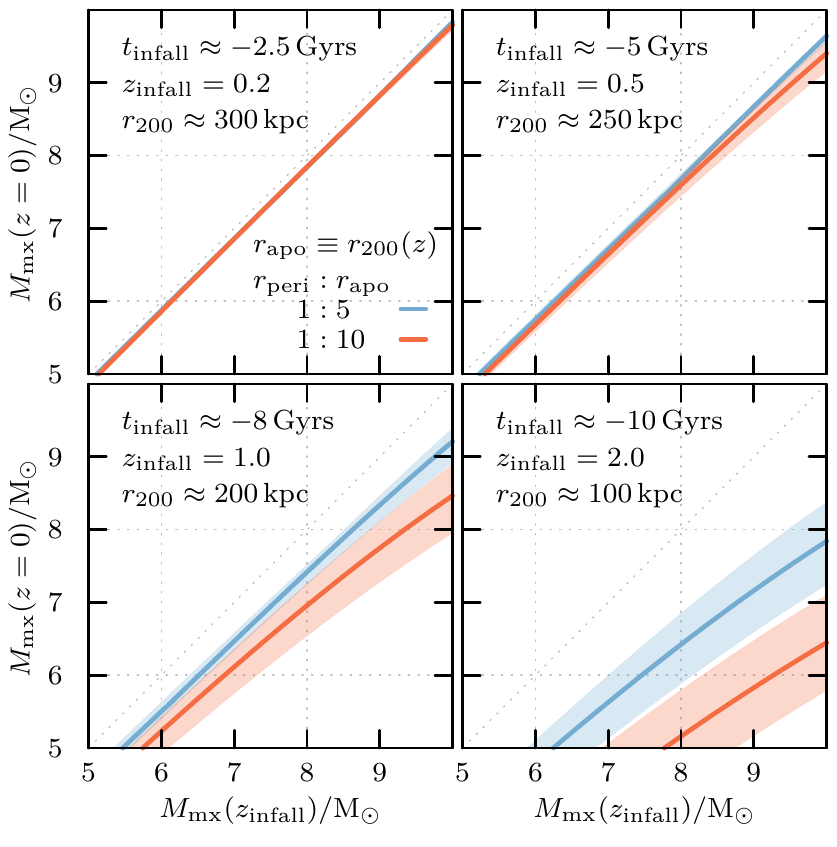}
\caption{Initial vs final bound mass of $\Lambda$CDM subhaloes accreted at different times into an isothermal halo with circular velocity $220\, \kms$. Panels show the bound mass of subhalos at $z=0$ as a function of their mass at infall, for four different infall redshifts ($z=0.2$, $0.5$, $1.0$, $2.0$), and orbital eccentricities of $1\rt5$ and $1\rt10$. Subhalos follow the mass-concentration relation at infall redshift, with shaded bands corresponding to $\pm0.1$ dex scatter in concentration. The apocentre of the subhalo orbit $r_\mathrm{apo}$ is chosen to correspond to the virial radius $r_{200}$ of the host halo at infall, as given in the legends. Note that subhalos of bound mass $10^8 < \Mmax / \Msol < 10^{10}$, accreted $10\,\Gyrs$ ago, have been stripped to less than $0.1$ per cent of their initial mass on orbits with $\rperi=20$ kpc. Many such remnants would have been artificially disrupted in direct cosmological simulations.}
 \label{fig:mmax_infall_vs_now}
\end{figure}

With this assumption, the virial radius of the host evolves ``passively'' from $r_{200}\sim 100$ kpc at $z=2$ to $\sim 300$ kpc at present \citep[``preudo-evolution'', see][]{Diemer2013}. Assuming that the apocentric distance of subhalos accreted at given redshift equals the host's current virial radius, Fig.~\ref{fig:mmax_infall_vs_now} shows the predicted masses at $z=0$ for subhalos accreted at $z=2$, $1$, $0.5$ and $0.2$. Two curves are shown, for $1\rt5$ (blue) and $1\rt10$ (red) pericentre-to-apocentre ratios, respectively. ``Error bands'' indicate the dispersion expected from the scatter in the mass-concentration relation ($\pm 0.1\,\mathrm{dex}$ in concentration).

In this illustration, most subhalos accreted at $z\sim 0.2$ (top-left panel in Fig.~\ref{fig:mmax_infall_vs_now}) have had time to complete at most one pericentric passage, and have therefore remained more or less unchanged since accretion. In contrast, subhalos with infall mass $\Mmax = 10^{10}\,\Msol$ accreted at $z=1$ have been stripped down to less than $\sim 10^9\,\Msol$, and those accreted at $z=2$ to less than $10^8\,\Msol$. 

The evolution of massive ($\Mmax \gtrsim 10^8\,\Msol$) subhalos that reach the inner regions of the Milky Way is of particular interest, as they could potentially host dwarf satellite galaxies that survive until the present. Fig~\ref{fig:mmax_vs_t} shows the evolution of subhalo mass $\Mmax$ as a function of time for subhalos with initial masses of $M\maxzero = 10^8\,\Msol$ and $M\maxzero = 10^{10}\,\Msol$, for fixed pericentre distances of $\rperi = 10\,\kpc$ and $20\,\kpc$. Most mass is lost within the first few Gyrs after accretion but even after $10$ Gyrs of evolution subhalos as massive as $10^{10}\,\Msol$ should leave behind remnants with $10^7$-$10^8\, M_\odot$ at pericentric distances of order $20$ kpc. These would be very poorly resolved---and maybe even missing--even in the best presently available cosmological hydrodynamical simulations, where the dark matter particle mass is typically of order $10^4 \sim 10^5\, M_\odot$ \citep{Onorbe2015Fire, SawalaFrenk2016, SchayeEagle2015}. This may have significant impact on $\Lambda$CDM predictions about the survival of faint satellites in the inner regions of the Milky Way, an issue that has attracted much interest in recent work, using cosmological simulations \citep[e.g.][]{Garrison-Kimmel2017,Richings2020}, controlled simulations \citep[e.g.][]{EPLG17, Sanders2018, vdBOgiya2018, EP2020} and semi-analytical approaches \citep[e.g.][]{Stref2019}.

\begin{figure}
 \centering
\includegraphics[width=8.5cm]{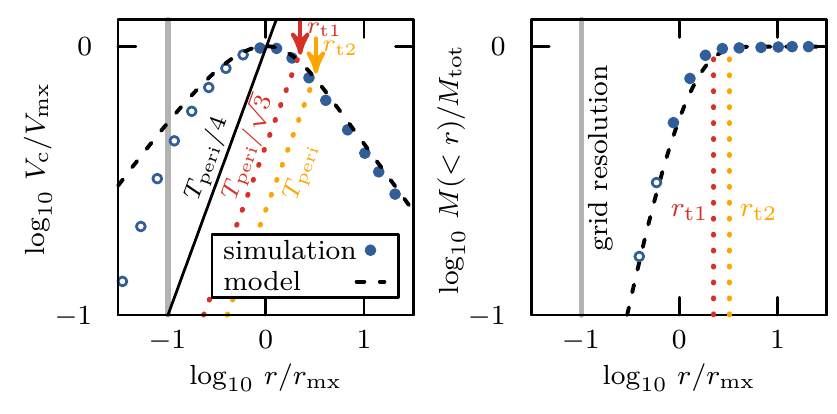}
 \caption{Circular velocity profile $V_\mathrm{c}$ (left panel) and enclosed mass $M(<r)$ (right panel) of a tidal remnant. Data from an $N$-body model (``simulation'' - the top-right snapshot in Fig.~\ref{fig:nbody_overview}) is shown using filled circles where unaffected by resolution, and using open circles where potentially affected by resolution. The analytical solution for a truncated NFW cusp (see Eq. \ref{eq:rhoasy}) reproducing the measured $\{\rmax,\vmax\}$ is shown using black dashed curves (``model'').
 In the left panel, the characteristic crossing time of $\tperi /4$ is shown using a solid black line, while the crossing times for two different simple definitions of tidal radii ($r_\mathrm{t1},r_\mathrm{t2}$) are shown using red and orange dashed lines. The same tidal radii are also marked in the right panel, showing that \emph{beyond} the tidal radius lies only a small fraction of the total bound mass $M_\mathrm{tot}$.  }
 \label{fig:tidal_radius}
\end{figure}

\begin{figure}
 \centering
\includegraphics[width=8.5cm]{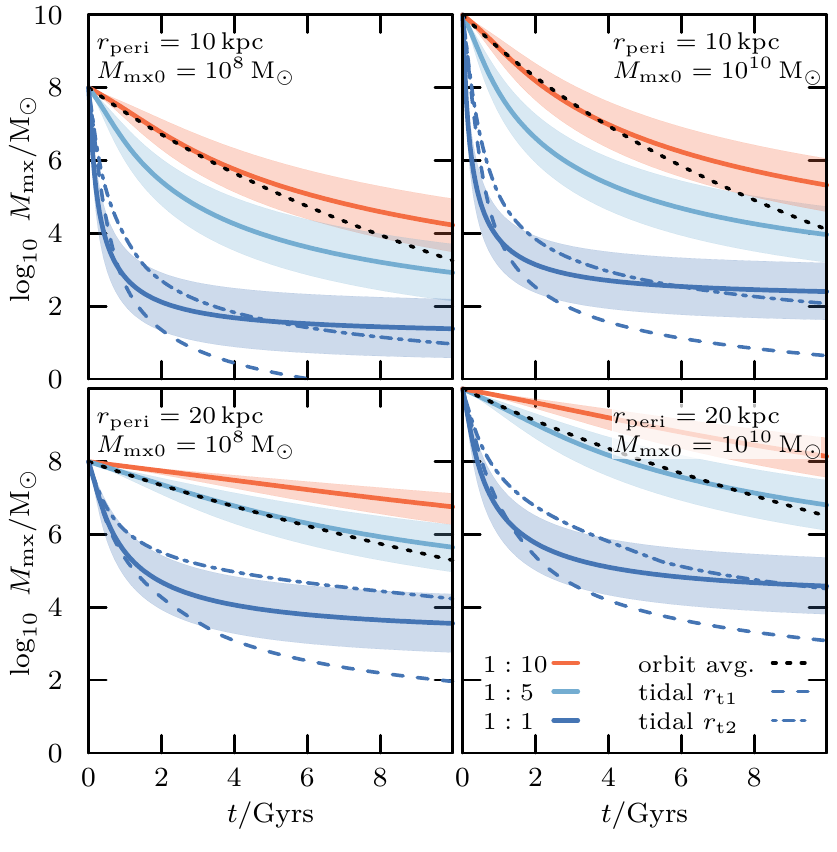}
 \caption{Evolution of bound mass on orbits of different eccentricity (Eqs.~\ref{EqYevol} and ~\ref{eq:model-regime-II-new}). Panels show subhalos with initial masses of $10^8\,\Msol$ (left column) and $10^{10}\,\Msol$ (right column) on orbits with pericentric distances of $10\,\kpc$ (top row) and $20\,\kpc$ (bottom row). The shaded bands correspond to $\pm0.1$ dex scatter in the initial ($z=0$) mass-concentration relation. Subhalos on near-circular orbits with orbital radii of $\sim10\,\kpc$ are stripped by several decades in mass over $10\,\Gyrs$. Black dotted curves (``orbit avg.'') show the orbit-averaged mass-evolution fitted to cosmological simulations (Eq.~\ref{eq:vdBmodel}, using parameters from \citealt{JiangVDB2016}), and blue dashed curves (``tidal $r_\mathrm{t1}$'' and ``tidal $r_\mathrm{t2}$'') show the mass-evolution on circular orbits as computed from simple tidal radius arguments (Eq.~\ref{eq:diffeq_tidal}), using the two tidal radii shown in Fig.~\ref{fig:tidal_radius}. }
 \label{fig:mmax_vs_t}
\end{figure}

\subsection{Comparison with previous work}
The tidal evolution of subhalos in cosmological and controlled simulations has been studied extensively in previous work, and this section aims to compare the predictions of the rate of tidal stripping of the model of Section~\ref{SecTimeEvol} to previous work.

\subsubsection{Comparison with orbit-averaged mass-loss rates}
The mass-loss rates of subhalos in cosmological simulations have been studied by \citet{vdBTormen2005}, who marginalize over all subhalo orbits, and propose a parametrization for the orbit-averaged mass loss rate of the form (here reproduced using the notation of \citealt{JiangVDB2016}):
\begin{equation}
\label{eq:vdBmodel}
 \diff m / \diff t = - \mathcal{A} ~ m ~(m/M)^\zeta  / \tau_\mathrm{dyn}  ~, 
\end{equation}
where $m$ is a measure of the subhalo mass, $M$ is a measure of the host halo mass, $\tau_\mathrm{dyn} $ is a measure for the crossing time of the host halo, and $\mathcal{A}$ and $\zeta$ are dimensionless constants.

For the parameters $\zeta$ and $\mathcal{A}$, \citet{JiangVDB2016} measure values of $\zeta \sim 0.07$ and $\mathcal{A} \sim 1.3$, consistent with the earlier findings of \citet{Giocoli2008}. Hence the average mass-loss as predicted by equation \ref{eq:vdBmodel} is close to exponential, and $m \rightarrow 0$ for $t \rightarrow \infty$. Using the tidal tracks of Section~\ref{sec:Tracks}, this also implies $\tmax \rightarrow 0$ for $t \rightarrow \infty$.

While a direct comparison of the \emph{orbit-averaged} mass-loss rates of Eq.~\ref{eq:vdBmodel} against the \emph{orbit-specific} rates of the model of Sec.~\ref{sec:general_model} is not straightforward, it is worth noting that the near exponential mass-loss described by Eq.~\ref{eq:vdBmodel} stands in stark contrast to the late-time behaviour described by the model of Eq.~\ref{EqYevol}, which predicts surviving remnants of non-zero mass, characterized by a crossing time of $\sim \tperi / 4$, set by the crossing time of the host halo at pericentre. With the sole purpose of showing the qualitative behaviour, Fig.~\ref{fig:mmax_vs_t} shows as a black dotted curve the mass-loss rate as predicted from integrating Eq.~\ref{eq:vdBmodel}, setting $\tau_\mathrm{dyn} = \tperi$.

\subsubsection{Comparison with tidal radius approaches}
Various authors have modelled the rate of mass-loss to be proportional to the mass \emph{outside} of some effective tidal radius $r_\mathrm{t}$  (\citealt{Taylor2001,Zentner2003,vdBTormen2005, Penarrubia2005, Diemand2007}, and more recently \citealt{vdb2018}), i.e.
\begin{equation}
\label{eq:diffeq_tidal}
 \diff m / \diff t = - \mathcal{B} ~m(>r_\mathrm{t}) / T_\mathrm{orb} 
\end{equation}
where $\mathcal{B}$ is a dimensionless constant. 
In this context, different recipes for the computation of the tidal radius have been proposed in the literature \citep[e.g.][]{Tormen1998, Klypin1999, Penarrubia2005, Read2006} and are reviewed in \citet{vdb2018}. To first order, these tidal radii $r_t$ are a measure for the region within the subhalo where the enclosed mean density is larger by some factor $\mathcal{C}$ than the enclosed mean density of the host halo at pericentre, e.g. 
\begin{equation}
 \bar \rho_\mathrm{sub}(< r_t) = \mathcal{C}~ \bar \rho_\mathrm{peri}  ~. 
 \label{eq:rtidaldef}
\end{equation}
In the following, we use $\mathcal{C}=3$ and $\mathcal{C}=1$ to define the two tidal radii $r_\mathrm{t1}$ and $r_\mathrm{t2}$, shown in Fig.~\ref{fig:tidal_radius}.
Using this simple definition of tidal radius, as well as the mass loss-dependent parametrization of the density profile discussed in Sec.~\ref{sec:Profile}, the mass evolution obtained from integrating Eq.~\ref{eq:diffeq_tidal} is plotted in Fig.~\ref{fig:mmax_vs_t} using blue-dashed curves. A value of $\mathcal{B}=6$ approximatively matches the initial mass evolution as computed from the model discussed in Sec.~\ref{SecTimeEvol}, and is consistent with the value measured by \citet{Diemand2007}. While the rate of mass loss decelerates as the remnant bound mass decreases, mass loss as described by the differential equation Eq.~\ref{eq:diffeq_tidal} in combination with the density profile evolution of Eq.~\ref{eq:rho_tidal} still eventually leads to fully disrupted tidal remnants, $m \rightarrow 0$ for $t \rightarrow \infty$. 

The reason for this asymptotic behaviour is easily understood by noting that for mass loss to stall ($\diff m / \diff t \rightarrow 0$), the differential equation Eq.~\ref{eq:diffeq_tidal} requires there to be no mass left outside the tidal radius, which is not met by the simple definitions of tidal radius of Eq.~\ref{eq:rtidaldef} in combination with the density profile parametrization of Sec.~\ref{sec:Profile}.
The mass loss model of Eq.~\ref{eq:diffeq_tidal} hence requires careful tailoring of the definition of tidal radius to the system in question, as discussed e.g. in \citet{vdb2018}. Specifically, to ensure a deceleration of tidal stripping that gives rise to a well-defined tidal remnant, a tidal radius definition is required which ensures that the mass beyond the tidal radius approaches zero sufficiently fast.

\subsection{Limitations of the model}
Several aspects of the parametrization for tidal stripping discussed in this work adopt simplifications that should be considered carefully when applying the model to physical systems.
\begin{itemize}
 \item[(i)] The rate of mass loss and the properties of the asymptotic remnant are set by the crossing time $\tperi$ of the host halo at pericentre. This is only well defined if one assumes $\tperi$ to be constant. This assumption is not valid for massive subhalos, as  dynamical friction would cause their orbits to decay, reducing their pericentric distances.
 \item[(ii)] Our models describe the rapid tidal evolution towards a remnant with a well-defined characteristic crossing time.
     Tidal remnants in our models are resolved with a small number of particles, $N(<\rmax) \gtrsim 3000$, and have characteristic radii that are only a few times the grid size of our finest spatial grid, $\rmax \gtrsim 8\, \Delta x$. These numerical limitations complicate the interpretation of the long-term evolution of our models, and prevent us from distinguishing clearly between an asymptotic timescale given by $T_{\rm asy}=0.22\,T_{\rm peri}$ (suggested by fits of Eq.~\ref{EqAsympt} to the combined results of all of our runs) and a slower ``secular'' evolution beyond this timescale. Assessing the long-term evolution of the tidal remnants using direct numerical simulations requires better numerical resolution than the one adopted in our work.
 \item[(iii)] Tidal stripping itself may cause changes to a subhalo's orbit because of asymmetries in the leading and trailing tidal stream and because of the self-gravity of the stream itself (see e.g. \citealp{White1983, HernquistWeinberg1989}, or more recently \citealp{Fuji2006,Fellhauer2007,Miller2020}). 
 \item[(iv)] Our simulations are based on a static host halo, without response to the gravity of the subhalo. While this setup seems well motivated for systems where the host mass enclosed within $\rperi$ is substantially larger than the mass of the subhalo, taking into account the host halo response will be important for mergers with larger host-to-subhalo mass ratios.
 \item[(v)] The host halo model used in this study is a singular isothermal sphere, with a circular velocity chosen to approximate the Milky Way potential (Eq.~\ref{EqHalo}). Tidal evolution in host halos with substantially different radial dependence on the tidal field may affect the numerical values proposed for the crossing time of the asymptotic remnant $T_\mathrm{asy}$, the asymptotic decay rate $\tau_\mathrm{asy}$, and the eccentricity `` delay'' factor $f_\mathrm{ecc}$. 
 \item[(vi)] All subhalo models considered were assumed to be collisionless, spherical, non-rotating, with an initially isotropic velocity dispersion.
 \item[(vii)] Our results apply to the accretion of single subhalos onto a smooth tidal field, and do not consider group infall: recent studies indicate that tidal stripping by \emph{clumpy} tidal fields may increase the rates of tidal stripping \citep{Stref2019,Delos2019}. 
\end{itemize}

\section{Summary and Conclusions}
\label{SecConc}

We have used N-body simulations of the tidal evolution of NFW halos in the potential of a much more massive host to investigate the time evolution of tidal mass loss, its dependence on orbital eccentricity and on the number of completed orbits, as well as the structural properties of the bound remnants. Our study also examines the effects of numerical limitations on the bound remnant structure, and the possibility that NFW subhalos almost always leave behind a self-bound remnant.

Some of these issues have been addressed by earlier work, but our conclusions clarify and extend some of the earlier conclusions, and shed light on the long-term survival of NFW remnants in the regime of heavy tidal mass loss. Our main conclusions may be summarized as follows.

The effect of tides on NFW subhalos leads to a self-bound remnant whose asymptotic properties are set solely by initial subhalo structure and the properties of the host halo at the orbital pericentre. We identify two regimes, depending on the ratio between the initial characteristic crossing time (density) of the subhalo, $T\maxzero$,  and the circular orbit timescale (density) of the host at pericentre, $\tperi$. Subhalos with $T\maxzero/\tperi< 2/3$ lose modest amounts of mass and approach asymptotically a remnant with a characteristic density set largely by its initial value.

On the other hand, subhalos with $T\maxzero/\tperi> 2/3$ lose large fractions of their initial mass and approach asymptotically a remnant whose characteristic timescale is set solely by the host density at pericentre; i.e., $\tasym \approx \tperi/4$ (Fig.~\ref{fig:tmax_time_october}). This result suggests that NFW subhalos are almost never fully disrupted, a result that may have important consequences on the long-term evolution and survival of luminous Milky Way satellites, as well as other implications for the studies of the distribution of dark matter on subgalactic scales.

As in earlier work, we find that the evolution of the characteristic parameters of the remnant (e.g., $\rmax$ and $\vmax$) depends solely on the total amount of mass lost, and that these parameters evolve along well-defined ``tidal tracks'', independent of orbital eccentricity or of the number of orbits required to strip the system (Fig.~\ref{fig:all_tracks}). Our improved numerical resolution allows us to extend and revise the tidal tracks proposed in earlier studies.

Numerical limitations lead poorly-resolved subhalos to deviate systematically from this track, making them more susceptible to tidal mass loss and possible full disruption. Such deviations may be used to identify remnants whose structure is not well converged numerically.  Finite spatial resolution (e.g., grid size or ``softening''; $\Delta x$), as well as time resolution (e.g., minimum timestep) impose obvious limits on the size or characteristic timescale of subhalos that may be resolved. For example, systems where $\rmax/\Delta x\lesssim8$ deviate from convergence and are prone to artificial disruption, regardless of the number of particles used.

In otherwise well-resolved systems, the number of particles used to resolve the subhalo places the ultimate constraint: all subhalos in our study start to deviate from convergence once they have been stripped to fewer than about $3000$ particles inside $\rmax$  (Fig.~\ref{fig:track_convergence}). This sets a high bar for the study of substructure in cosmological N-body simulations.

The shape of the mass profile of a tidally stripped subhalo deviates from the initial NFW shape, and is well described by an exponentially-truncated NFW density profile (Eq.~\ref{eq:rho_tidal}). The truncation ``radius'' is set solely by the mass fraction that remains bound to the remnant. All heavily stripped NFW subhalos thus converge asymptotically to the same mass profile shape, an exponentially truncated NFW cusp (Fig.~\ref{fig:profile_evolution}).

The time evolution of the structural parameters of a subhalo may be well approximated by a simple function (Eq.~\ref{EqYevol}) with a few scaling parameters that are well constrained by our simulation results. The main effect of orbital eccentricity is to ``delay'' the evolution relative to subhalos on circular orbits at equal pericentre. The delay factor, $f_{\rm ecc}$, is also well constrained by our simulation results (Eq.~\ref{EqFecc}).

Our results thus provide a full description of the tidal evolution of NFW subhalos, with the caveat that these results apply to the regime where the orbits have well-defined pericentric distances (i.e., the potential is approximately spherical and orbits are unaffected by tidal loss or dynamical friction) and the host potential does not evolve substantially with time. Although these caveats imply that our results cannot be used to make direct predictions for the properties of substructure in a $\Lambda$CDM halo, they can be used to interpret the results of cosmological simulations, and to identify their deficiencies and/or limitations.  Our results may also be combined with cosmological simulations to place constraints on the abundance and structure of surviving subhalos and on their relation with ultra-faint satellites and other dark matter-bound structures in the inner regions of the Galaxy. We plan to apply the lessons learned here to a number of pressing questions concerning substructure in CDM halos in future contributions.

\section*{Acknowledgements}
RE wants to thank J. Pe\~narrubia for discussions which were at the base of shaping ideas behind this work. We acknowledge useful discussions with Laura Sales, and thank the anonymous referee for detailed comments.
RE also acknowledges support provided by a CITA National Fellowship and by funding from the European Research Council (ERC) under the European Unions Horizon 2020 research and innovation programme (grant agreement No. 834148). 
This work used the DiRAC@Durham facility managed by the Institute for Computational Cosmology on behalf of the STFC DiRAC HPC Facility (\url{www.dirac.ac.uk}). The equipment was funded by BEIS capital funding via STFC capital grants ST/K00042X/1, ST/P002293/1, ST/R002371/1 and ST/S002502/1, Durham University and STFC operations grant ST/R000832/1.

\section*{Data availability}
The data underlying this article will be shared on reasonable request to the corresponding author.

\footnotesize{
\bibliography{tidal_remnants}
}

\appendix
\section{Numerical convergence}
\label{SecNumConv}
\label{sec:AppendixA}

Numerical resolution imposes strong limits on the ability of simulations to follow the tidal evolution of subhalos. Most critical are the finite timestepping, spatial resolution, and number of particles used in a simulation. We explore in this Appendix the impact of such limitations and the constraints they place on our results.

As stated in Sec.~\ref{SecNumRes}, our simulations evolve subhalos with a single, constant timestep set to $\Delta t=0.025\times \min(T\maxzero,\tperi)$. This timestep is shown as a dashed diagonal line in Fig.~\ref{fig:vc_curves} ($\times 10$ to fit in the figure) and is clearly much shorter than the subhalo crossing time at a radius equal to the best grid spatial resolution, $\sim r\maxzero/128$. Fixing the timestep this way reduces the dimensionality of the problem, leaving only the spatial (grid) resolution and the number of particles for us to consider.

To do so, we perform two series of simulations: one where we fix the number of subhalo particles to $N=10^7$, the maximum in our runs, and vary the grid size systematically from $\Delta x\approx r\maxzero/128$ to $r\maxzero/32$; and another where we fix $\Delta x $ to $r\maxzero/128$ and vary the number of particles from $10^7$ to $10^5$. We choose for these tests subhalos on $1\rt5$ eccentric orbits ($\rperi = 40\,\kpc$) with initial crossing times $0.5 < T\maxzero/\tperi <2$.

\begin{figure}
 \centering
\includegraphics[width=8.5cm]{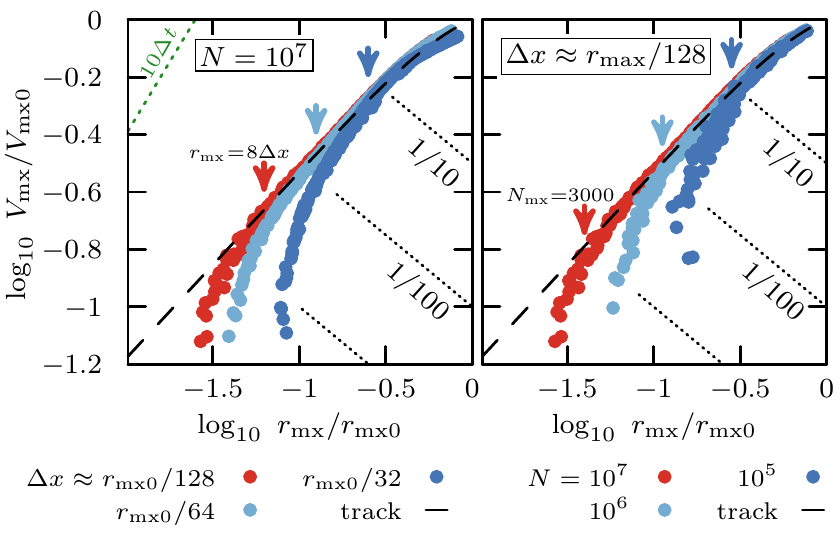}
\caption{{\it Left:} Evolution of $\rmax$ and $\vmax$ for $10^7$-particle halos run with three different grid sizes for the highest resolution mesh. Note that systems start to deviate from the tidal track (shown with a dashed black curve, Eq.~\ref{eq:tracks}) when the characteristic radius of the remnant approaches $\rmax\approx 8\,\Delta x$. {\it Right:} same as left, but for a series of runs with fixed time and spatial resolution, but varying the number of particles of the initial halo. Note that remnants artificially deviate systematically from the tidal track when the remnant is resolved with fewer than $\sim 3000$ particles within $\rmax$. } 
 \label{fig:track_convergence}
\end{figure}

The ``tidal tracks'' that result are shown in  Fig.~\ref{fig:track_convergence}. Each symbol corresponds to parameters measured at a successive apocentric passage, normalized to the initial values. The left panel shows the effect of varying the grid size. As the spatial resolution deteriorates, subhalos deviate systematically from the converged tidal track (indicated by the dashed black curve) toward longer crossing times and lower characteristic densities. The arrows indicate the radius corresponding to $\rmax = 8 \Delta x$, which, in each case, is a good diagnostic of the minimum ``size'' a subhalo must have for its characteristic parameters to be properly resolved. More precisely, subhalos with $\rmax < 8 \Delta x$ have characteristic timescales, $\tmax$, that deviate more than $10$ per cent from the timescale expected from the tidal track.

The right-hand panel in Fig.~\ref{fig:track_convergence} is analogous to the one on the left, but for the series of runs where the number of particles is varied. The arrows in this case indicate the location of subhalos where the number of particles inside $\rmax$, $ N_\mathrm{mx}=\Mmax/m_{\rm p}$, drops below $\sim3000$ (here $m_p$ is the mass per particle). This simple criterion again identifies the minimum number of particles needed to resolve the characteristic parameters of a tidally-affected NFW subhalo, in the sense than tidal remnants with $N_\mathrm{mx}<3000$ typically have crossing times that deviate by more than $10$ per cent from the converged tidal track. The analysis throughout the paper is based on results obtained for subhalos that satisfy simultaneously both criteria (i.e., $\rmax > 8 \Delta x$ and $N_\mathrm{mx}>3000$).

For our simulations with $N=10^7$ and $\Delta x \approx r\maxzero/128$, resolution is maily limited by the grid size, and the condition $\rmax > 8 \Delta x$ implies that numerical limitations begin to dominate once $\Mmax$ has been reduced to about $0.3$ per cent of its initial value. This sets the limits of the most highly stripped system effectively probed by our simulations: $\Mmax/M\maxzero \approx 1/300$; or $\vmax/V\maxzero \approx 1/5$; or $\rmax/r\maxzero\approx 1/16$; or $\tmax/T\maxzero\approx 1/3$.

\label{lastpage}
\end{document}